\documentclass[fleqn,usenatbib]{mnras}

\usepackage[T1]{fontenc}

\DeclareRobustCommand{\VAN}[3]{#2}
\let\VANthebibliography\thebibliography
\def\thebibliography{\DeclareRobustCommand{\VAN}[3]{##3}\VANthebibliography}

\usepackage{graphicx}	% Including figure files
\usepackage{amsmath}	% Advanced maths commands
\usepackage[dvipsnames]{xcolor}
\usepackage{amsmath}  % more options for equations (matrix environment, etc.)
\usepackage{amssymb}  % fancy merger rate symbol, etc.
\usepackage{upgreek}  % for upright Greek symbols (\uppi, \upmu)
\usepackage{afterpage}  % individual page for figures
\usepackage{times}  % make font much more compact
\usepackage{paralist}  % compact lists (compactitem)
\usepackage{pbox}  % line break inside table cell
\usepackage{booktabs}  % vertical spaces in table
\usepackage[flushleft]{threeparttable}  % tables with notes, etc.
\usepackage{array}
\usepackage{mathtools}  % fancy matrices
\usepackage{subfigure}
\usepackage{grffile}

\newcommand{\Msun}{{\rm M}_{\odot}}

\title[The ICL in simulations and observations]{Photometric analysis of the intracluster light in the TNG300 simulation and wide-field observations}

\author[D. Montenegro-Taborda et al.]
{
	\parbox{18cm}{
	  Daniel Montenegro-Taborda,$^{1}$\thanks{E-mail: d.montenegro@irya.unam.mx}
	  Vicente Rodriguez-Gomez,$^{1}$
    Vladimir Avila-Reese,$^{2}$
    Bernardo Cervantes-Sodi,$^{1}$
	  Matthias Kluge,$^{3}$
    Aditya Manuwal,$^{2}$
    Annalisa Pillepich$^{4}$ and
	  Lars Hernquist$^{5}$
	}
	\vspace{0.3cm} \\ 
	$^{1}$ Instituto de Radioastronom\'ia y Astrof\'isica, Universidad Nacional Aut\'onoma de M\'exico, A.P. 72-3, 58089 Morelia, Mexico \\
	$^{2}$ Instituto de Astronom\'ia, Universidad Nacional Aut\'onoma de M\'exico, A.P. 70-264, 04510 CDMX, Mexico \\
    $^{3}$ Max Planck Institute for Extraterrestrial Physics, Giessenbachstrasse 1, 85748 Garching, Germany \\
    $^{4}$ Max-Planck-Institut f\"ur Astronomie, K\"onigstuhl 17, D-69117 Heidelberg, Germany \\
    $^{5}$ Harvard-Smithsonian Center for Astrophysics, 60 Garden Street, Cambridge, MA 02138, USA \\
}

\date{Accepted XXX. Received YYY; in original form ZZZ}

\pubyear{2025}

\begin{document}
\label{firstpage}
\pagerange{\pageref{firstpage}--\pageref{lastpage}}
\maketitle

\begin{abstract}
We present a robust, apples-to-apples comparison between the photometric properties of the intracluster light (ICL) in the TNG300 magnetohydrodynamic cosmological simulation and those in Wendelstein Wide Field Imager (WWFI) observations. This is accomplished by generating synthetic $g'$-band images of 40 massive ($\log\left(M_{\rm 200, crit}/{\rm M}_{\odot}\right) > 14.5$) TNG300 clusters at $z \approx 0.06$, closely mimicking WWFI observations, and then performing identical photometric calculations on the synthetic and real images. Importantly, we apply the same observationally motivated satellite-masking procedure to both data-sets, which effectively removes any possible biases introduced by the halo finder. We first analyze the light distribution of the `smooth' stellar component of each cluster, composed of the brightest cluster galaxy (BCG) plus the ICL, and find that it tends to be about twice as extended in TNG300 than in observations, while also being approximately 1 $g'$ mag arcsec$^{-2}$ brighter. We then quantify $f_{\rm ICL}$, the ICL fraction relative to the BCG+ICL, by considering several ICL definitions: (i) the light dimmer than a surface brightness cut at 27 $g'$ mag arcsec$^{-2}$, (ii) the excess light over a de Vaucouleurs profile, (iii) the light beyond twice the half-light radius ($2 r_{\rm half}$), and (iv) the light beyond a fixed circular aperture of 30, 50, or 100 kpc. For most definitions, the median $f_{\rm ICL}$ is consistent between simulation and observations. However, the observations exhibit larger scatter in $f_{\rm ICL}$, which we attribute primarily to observational uncertainties in the total BCG+ICL luminosity rather than `true' cluster-to-cluster variation in the real Universe. We also find that most methods yield median $f_{\rm ICL}$ values near 0.3, which is consistent with a BCG/ICL transition radius around $2 r_{\rm half}$.
\end{abstract}


%%%%%%%%%%%%%%%%%%%%%%%%%%%%%%%%%%%%%%%%%%%%%%%%%%

%%%%%%%%%%%%%%%%% BODY OF PAPER %%%%%%%%%%%%%%%%%%

\section{Introduction}

The luminous components within galaxy clusters are intricate and diverse, reflecting the complex nature of their formation process. Among these components is the intracluster light (ICL), a diffuse collection of stars that surrounds the brightest cluster galaxy (BCG) and that is not gravitationally bound to any individual galaxy but rather to the cluster as a whole \citep{zwicky1951_ICL}. The ICL consists primarily of stars that have been accreted through mergers or stripped from galaxies during their interactions within the cluster environment \citep[][]{Gallagher1972, Ostriker1975, White1976, Hausman1978}. As a result, the ICL serves as a fossil record of the formation and evolutionary history of galaxy clusters \citep[see][for reviews]{contini2021_Review, montes2022Nature}. Additionally, the ICL can serve as a valuable tracer for inferring the distribution of dark matter (DM) throughout the cluster halo. This possibility has been explored and supported in previous studies \citep[e.g.][]{Montes&Trujillo2019, sampaio-santos2021, Diego2023, Yoo2024}.

Despite the astrophysical importance of the ICL, quantifying the amount of light associated with it is a topic of considerable debate in the literature, both observationally and theoretically \citep[e.g.][]{rudick2011, Tang2018, Pillepich2018a, contini2021_Review, kluge2021photometric, montes2022Nature, Brough+2024}. The uncertainty arises because the BCG and ICL are highly intermixed because of their similar growth channels, making it difficult to separate them using conventional methods such as fixed radial cuts or surface brightness (SB) thresholds. Furthermore, simulations show that the stellar mass budgets of both the BCG and ICL are dominated by \textit{ex situ} (i.e. accreted) stars \citep{montenegro2023, montenegro2025}, which raises the question of whether the BCG and the ICL should be considered as separate entities from a theoretical standpoint, even if provided with full information about the origin of every stellar particle.

% \citep{cooper2015, Rodriguez-Gomez2016, Pillepich2018a, pulsoni2021, Remus2022, chun2023, tang2023,montenegro2023, brown2024, montenegro2025}

Observationally, a wide range of methods have been proposed in the literature to identify the ICL, of which the following stand out: fixed radial cuts \citep[e.g.][]{Kravtsov2018}, SB thresholds \citep[e.g.][]{feldmeier2004, rudick2011, montes2018, furnell2021, kluge2021photometric, martinez-lombilla2023}, SB profile fitting \citep[e.g.][]{gonzalez2005, zibetti2005, kluge2021photometric, ahad2023}, multi-galaxy fitting \citep[e.g, ][]{morishita2017}, wavelet decomposition \citep[e.g.][]{darocha2005, ellien2021}, and kinematic analyses using luminous tracers such as planetary nebulae \citep[e.g.][]{arnaboldi2022} or globular clusters \citep[e.g.][]{alamo2017}.

On the simulation side, common methods to identify the ICL are based on the gravitational potential as well as velocity distribution fitting techniques \citep[][]{dolag2010, rudick2011, canas2020stellar, proctor2024}. However, since these methods rely on simulation-only information such as gravitational binding energies and 3D velocities, which are not directly accessible observationally, most simulation studies adopt ICL definitions that are conceptually similar to those used by observers, such as fixed radial cuts \citep[e.g.][]{Pillepich2018a, Henden2020, montenegro2023, montenegro2025}, SB thresholds \citep[e.g.][]{rudick2011, cui2014, cooper2015, Brough+2024}, and SB profile fitting \citep[e.g.][]{cooper2015, remus2017, Brough+2024}.

Even for theoretical and observational studies that adopt the `same' ICL definition, another key source of uncertainty arises from the direct comparison of simulation outputs with observational results. In particular, when quantifying the ICL fraction -- the fraction of light or stellar mass in the ICL relative to the BCG+ICL or to the whole cluster -- in simulations and observations, most comparisons are carried out at face value, ignoring the observational and numerical biases inherent to each method. Some factors affecting such comparisons between simulations and observations include limitations posed by projected structures \citep{CastroRodriguez2009, demaio2015} and instrumental sensitivity \citep{Tang2018}, which restricts the sample size and completeness of observational studies.

An additional factor affecting such comparisons is how satellites are separated from the diffuse stellar background. In simulations, there is a wide variety of `halo finders' that can be used for such a task, which can operate quite differently in how they separate substructures, especially within dense cluster environments \citep{muldrew2011, han2012,onions2012, Behroozi2015}. Notably, properties measured near the outskirts of subhaloes present the largest discrepancies, highlighting the impact of the halo finder selection. For example, \cite{canas2020stellar} found that the masses of simulated satellite galaxies are larger when they are identified with a phase-space (6D) halo finder (\textsc{velociraptor}; \citealt{canas2019-galaxy-finder}) than when they are identified with a density-based structure finder (such as \textsc{subfind}; \citealt{Springel2001-SUBFIND}), which has a direct impact on the amount of stellar mass associated to the BCG+ICL. 

A more reliable approach for carrying out comparisons between simulations and observations, which almost completely eliminates any biases due to the halo finder and projection effects, involves `forward modeling' of the simulation data, such as generating synthetic images of simulated clusters and then analyzing these images using observational methods. This `apples-to-apples' approach has been successfully applied in topics such as galaxy morphology \citep{Rodriguez-Gomez2019}, stellar haloes \citep{merritt2020, ardila2021} and the ICL \citep{cooper2015, Tang2018}, among others. Such forward-modeling techniques have gained traction in recent years thanks to increasingly large and realistic hydrodynamic cosmological simulations, such as the one we use in this paper, allowing detailed studies of the ICL across a wide range of cluster masses \citep[][]{Pillepich2018a}.

A compelling analysis of the ICL in simulations using observationally motivated techniques was recently carried out by \cite{tang2023}, who generated synthetic images \citep{Tang2018} and then applied a novel surface brightness level segmentation procedure \citep[SBLSP,][]{tang2020, tang2021} to study the properties of the diffuse light in groups and clusters from the TNG100 simulation. They found similarities between the age, metallicity, and colour of the diffuse light and those of satellite galaxies, lending support to the idea that the diffuse light in groups and clusters originates from satellites (i.e. \textit{ex situ} processes).

In this study, we present a rigorous, `apples-to-apples' comparison between the photometric properties of galaxy clusters in the TNG300 simulation \citep{Marinacci2018, Naiman2018first, Nelson2018, Pillepich2018a, springel2018} and in observational data \citep{Kluge+2020}. To this end, we generate and analyze synthetic $g'$-band images of 40 massive TNG300 clusters at $z \approx 0.06$, such that they are directly comparable to the images of 170 clusters presented by \cite{Kluge+2020}. Our analysis includes photometric determinations of BCG+ICL sizes and SB profiles, as well as ICL fractions, for which we adopt definitions similar to those used by \cite{kluge2021photometric}. 

This paper is structured as follows. In Section \ref{sec:methods} we describe the observational and simulation data products, as well as the methodology used to generate and process the synthetic images. In Section \ref{sec:bcg+icl}, we study properties of the BCG+ICL as a single entity, focusing on BCG+ICL sizes and SB profiles. Section \ref{sec:icl_fraction} presents our main results on the ICL fraction in simulation and observations. Finally, we summarize and discuss our results in Section \ref{sec:summary_and_discussion}.

\section{Methods}
\label{sec:methods}

In this section we describe the observational and simulated data sets used throughout this work, as well as the data processing steps necessary to ensure a fair comparison between observations and simulations.

\subsection{Observations}
\label{subsec:observations}

We use the image sample presented by \cite{Kluge+2020}, which consists of $g'$-band\footnote{The $g'$ filter is a modified version of the original $g$ filter from the Sloan Digital Sky Survey, offering increased sensitivity at bluer wavelengths and a more uniform transmission curve over its entire wavelength range, which spans approximately from 3900 to 5700 \AA.} observations of 170 clusters, mostly in the northern hemisphere, obtained by the Wendelstein Wide Field Imager (WWFI) at the Wendelstein Observatory, Germany. The observed clusters have a mean redshift $z = 0.059$ with standard deviation $\Delta z = 0.022$. This represents a fractional variation in $(1 + z)$ of about 2 per cent, or a fractional variation in $(1+z)^4$ (the SB dimming correction) of about 10 per cent. Thus, since we do not expect a large amount of scatter in our results to be driven by SB dimming and other cosmological effects, we will work with fluxes in the observer's frame (instead of converting them to the rest frame of the source).

The observed clusters are mostly from the Abell--Corwin--Olowin cluster sample \citep{Abell1989} and have gravitational masses typically in the range $\log (M_{\rm g}/\Msun) = 14.5$--$15$, although only 38 of the 170 objects have gravitational mass measurements \citep{kluge2021photometric}. We refer to this subsample of 38 objects as the \textit{main} WWFI sample, which will be used for most comparisons to simulations. The stacked images have a pixel scale of 0.2 arcsec, a Gaussian point spread function (PSF) with a median full width at half-maximum (FWHM) of 1.2 arcsec, a median exposure time of 3120 s, and a limiting surface brightness (SB) around 30 $g'$ mag arcsec$^{-2}$, making them ideal for deep photometric studies of the ICL. Some examples of observed clusters are shown in Appendix \ref{app:cluster_images}, along with their corresponding satellite masks.

\subsection{The TNG300 simulation}
\label{subsec:simulation}

We employ the TNG300 simulation of the IllustrisTNG Project \citep{Marinacci2018, Naiman2018first, Nelson2018, nelson2019illustristng, Pillepich2018a, springel2018}, which follows the evolution of roughly $2 \times 2500^3$ resolution elements within a cubic volume of approximately 300 Mpc per side. The masses of the dark matter (DM) and baryonic resolution elements in TNG300 are $m_{\rm DM} \approx 5.9 \times 10^7 \, \Msun$ and $m_{\rm b} \approx 1.1 \times 10^7 \, \Msun$, respectively. The gravitational softening length of stars and DM particles at $z < 1$ is kept fixed at $\epsilon_{\ast} = 1.48$ kpc. The galaxy formation model is described in detail in \cite{Pillepich2018} and was designed to approximately reproduce several observables at $z=0$, including the galaxy stellar mass function, the stellar-to-halo and black hole-to-halo mass relations, halo gas fractions, and galaxy sizes, as well as the global star formation rate density since $z=8$. The cosmological parameters used in the simulations are consistent with Planck 2015 measurements \citep{ade2016planck}: $\Omega_\textrm{m}$ = 0.3089, $\Omega_{\Lambda}$ = 0.6911, $\Omega_\textrm{b}$ = 0.0486, $h$ = 0.6774, $\sigma_8$ = 0.8159, and $n_\textrm{s}$ = 0.9667.

The main output of the TNG300 simulation consists of 100 snapshots between $z=20$ (snapshot 0) and $z=0$ (snapshot 99). Haloes are identified for each snapshot using the friends-of-friends algorithm \citep{davis1985evolution}, while gravitationally bound substructures within them are found with \textsc{subfind} \citep{Springel2001-SUBFIND, dolag2009substructures}. The centre of each halo corresponds to the most bound particle of the central or `background' subhalo.

\begin{figure}
  \centering
  \includegraphics[width=8.5cm]{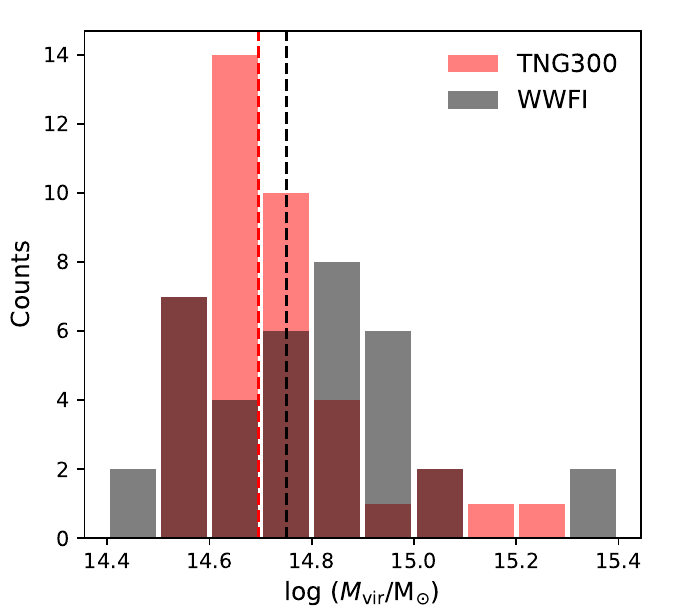}
	\caption{Histogram showing the virial masses of the simulated (red) and observed (black) galaxy cluster samples at $z = 0.0585$ and $\bar{z} = 0.059$, respectively. The vertical dashed lines indicate the corresponding medians. Note that \protect\cite{kluge2021photometric} report masses for only 38 out of the 170 clusters.}
	\label{fig:mvir}
\end{figure}

In order to roughly match the observational sample from \cite{Kluge+2020}, we select all TNG300 clusters at $z = 0.0585$ (snapshot 94) with halo masses $\log \left(M_{\rm 200, crit} / \Msun \right) \geq 14.5$, where $M_{\rm 200, crit}$ is the total mass within $r_{\rm 200, crit}$, the radius enclosing an average density equal to 200 times the critical density of the Universe, resulting in a total of 40 systems. Fig. \ref{fig:mvir} compares the virial masses ($M_{\rm vir}$) of the simulated clusters (computed using the spherical collapse approximation of \citealt{bryan1998}) against the gravitational masses of the main WWFI sample (see Section \ref{subsec:observations}), showing that both samples span a similar mass range. For simplicity, we will refer to the gravitational masses from \cite{kluge2021photometric} as virial masses. Although these two mass definitions are technically not the same, we do not expect large systematic differences between them.

\subsection{Synthetic images}
\label{subsec:synthetic_images}

\begin{figure*}
  \centering
  \includegraphics[width=\linewidth]{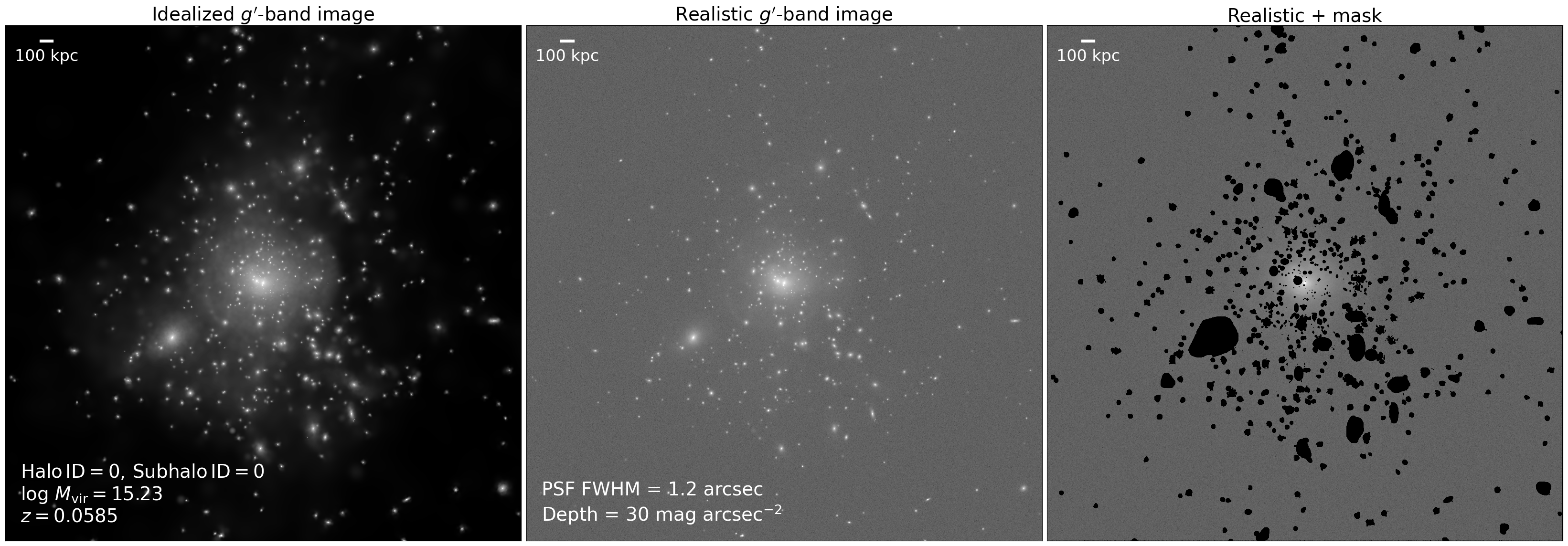}
	\caption{Idealized (left), realistic (middle), and masked (right) $g'$-band synthetic images of the most massive cluster in the TNG300 simulation at $z = 0.0585$, as described in Sections \ref{subsec:synthetic_images} and \ref{subsec:image_masking}. The field of view is $2 r_{\rm 200, crit}$.}
	\label{fig:cluster_multipanel}
\end{figure*}

For each of the 40 simulated clusters, we create dustless\footnote{We do not expect dust attenuation to play an important role in our analysis, since both the BCG and ICL have very low dust densities. While dust could have a non-negligible effect on the luminosities of satellite galaxies, these are entirely removed from our analysis, as explained in Section \ref{subsec:image_masking}.} $g'$-band images following the methodology described in \cite{Rodriguez-Gomez2019}. Briefly, each stellar particle is treated as a simple stellar population (SSP) located at $z = 0.0585$, and its $g'$-band flux \textit{in the observer's frame} is calculated using \textsc{GALAXEV} models \citep{bruzual2003}. The flux units are converted to analog-to-digital units (ADU) per second with a photometric zero-point (ZP) of 30 $g'$ mag. The light from each SSP is smoothed by means of an adaptive kernel with a smoothing scale equal to the distance to its 64th nearest neighbour.\footnote{This slightly differs from the original implementation in \cite{Rodriguez-Gomez2019}, where the distance to the 32nd nearest neighbour was used. Our results are not overly sensitive to the amount of adaptive smoothing applied to the synthetic images, as discussed in Appendix \ref{app:image_smoothing}.} Finally, the smoothed light contributions from all the individual SSPs are added within a square region of $2 \times r_{\rm 200, crit}$ per side, with the BCG -- the central subhalo according to \textsc{subfind} -- at the centre by construction. The pixel scale is set to 0.2 arcsec in order to match the WWFI observations, which corresponds to 0.234 kpc at $z = 0.0585$. All images are projected along the $z$-direction of the box, which effectively represents a random orientation for each simulated cluster.

The above procedure results in so-called \textit{idealized} synthetic images, as would be hypothetically observed by a `perfect' instrument with a point-like PSF and an infinite signal-to-noise ratio (S/N). The left-hand panel of Fig. \ref{fig:cluster_multipanel} shows an example of an idealized image for the largest cluster in the TNG300 simulation.

The next step is to apply some `realism' to the idealized synthetic images, which we do in three stages:
\begin{itemize}
    \item[(i)] PSF convolution. We convolve the images with a Gaussian PSF with FWHM = 1.2 arcsec, which corresponds to the median PSF FWHM of the observations (Section \ref{subsec:observations}). \\

    \item[(ii)] Shot noise. According to \citet{kosyra2014}, a source of 25.4 $g'$ mag produces an instrumental flux of 1 e$^-$ s$^{-1}$ in the detector. Since the synthetic images have units of ADU s$^{-1}$ with ZP = 30 $g'$ mag, this represents a gain of 0.0145 e$^-$ ADU$^{-1}$. Furthermore, since the median exposure time of the observations is 3120 s, we obtain an \textit{effective} gain of 45.1 e$^-$ (ADU/s)$^{-1}$, i.e. the multiplicative factor used to convert the image units from ADU s$^{-1}$ to electrons. We then replace each pixel value with a sample drawn from a Poisson distribution with an average equal to the expected number of electrons in that pixel, and finally return the image to its original units of ADU s$^{-1}$. \\

    \item[(iii)] Background noise. We add to each pixel a value drawn from a Gaussian distribution with standard deviation $\sigma_{\rm sky} = $ 2 ADU s$^{-1}$. This value of $\sigma_{\rm sky}$ was obtained by applying sigma-clipping within circular annuli to the real images, selecting the minimum-sigma annulus for each image, and then taking the median of all the minimum sigma values. This results in a representative, realistic value for the standard deviation of the background noise.
    
\end{itemize}
The middle panel of Fig. \ref{fig:cluster_multipanel} shows the `realistic' version of the idealized synthetic image shown on the left-hand panel. Further examples are shown in Appendix \ref{app:cluster_images}.

\subsection{Image masking}
\label{subsec:image_masking}

An important distinction between the current work and other simulation-based studies of the ICL is that we do not rely on the halo finder (\textsc{subfind}) to separate satellites from the BCG+ICL in the simulation. Instead, we choose to mask satellites directly on the synthetic images in an observationally motivated fashion by closely following the procedure described in \cite{Kluge+2020}, which we attempt to summarize as follows.

First, a rough model of the BCG+ICL is created by estimating the sigma-clipped median in $51 \times 51$ square apertures and interpolating the resulting grid of points. We use the \texttt{Background2D} class of the \texttt{photutils} package \citep{bradley2021_photutils} for this task. The BCG+ICL model is then subtracted from the image. A satellite mask is generated for the residual by combining a small and a medium mask, which are created by convolving the residual with 2D Gaussian filters with $\sigma = $ 5 px and $\sigma = $ 21 px, respectively, and then masking all pixels with values satisfying \citep[][their eq. 8]{Kluge+2020}
\begin{equation}
    T(x, y) \geq T_0 \cdot {\rm rms}(x, y) \cdot \left(\frac{\sqrt{{\rm rms}(x, y)}}{{\rm median(\sqrt{{\rm rms}(x, y)})}} \right)^{-1},
\end{equation}
where the threshold $T_0 = 0.3$ is given in units of the local S/N per pixel.\footnote{\cite{Kluge+2020} used $T_0 = 0.15$.} The small and medium masks are then expanded by convolving them with circular kernels of radii $r = $ 9 px and $r = $ 11 px, respectively, and combined into a final candidate mask. These steps represent the automatic stage of the mask creation procedure.

Finally, in order to correct for artifacts near the central regions of the BCG+ICL (since the initial $51 \times 51$ filter is too coarse to properly model highly concentrated profiles), these regions are unmasked and remasked by hand \citep{Kluge+2020}. Similarly, if there are other bright and extended regions in the image, excluding the BCG+ICL of interest, the masks in these regions are manually expanded to where the $\sim$29--30 $g'$ mag isophotes are expected to be. The right-hand panel of Fig. \ref{fig:cluster_multipanel} shows the realistic image from the middle panel after applying the satellite mask. Further examples of masked images are shown in Appendix \ref{app:cluster_images}.

We note that the BCG was redefined for 2 of the 40 simulated systems during this stage, since the ICL was visually determined to be centred around a bright galaxy other than the central originally determined by \textsc{subfind}. This again represents a more observationally motivated approach to studying galaxy clusters, since the BCG is visually confirmed and does not necessarily correspond to the central object determined by the halo finder.

\subsection{Background subtraction}
\label{subsec:background_subtraction}

\begin{figure*}
  \centering
  \centerline{\hbox{
    \includegraphics[width=9cm]{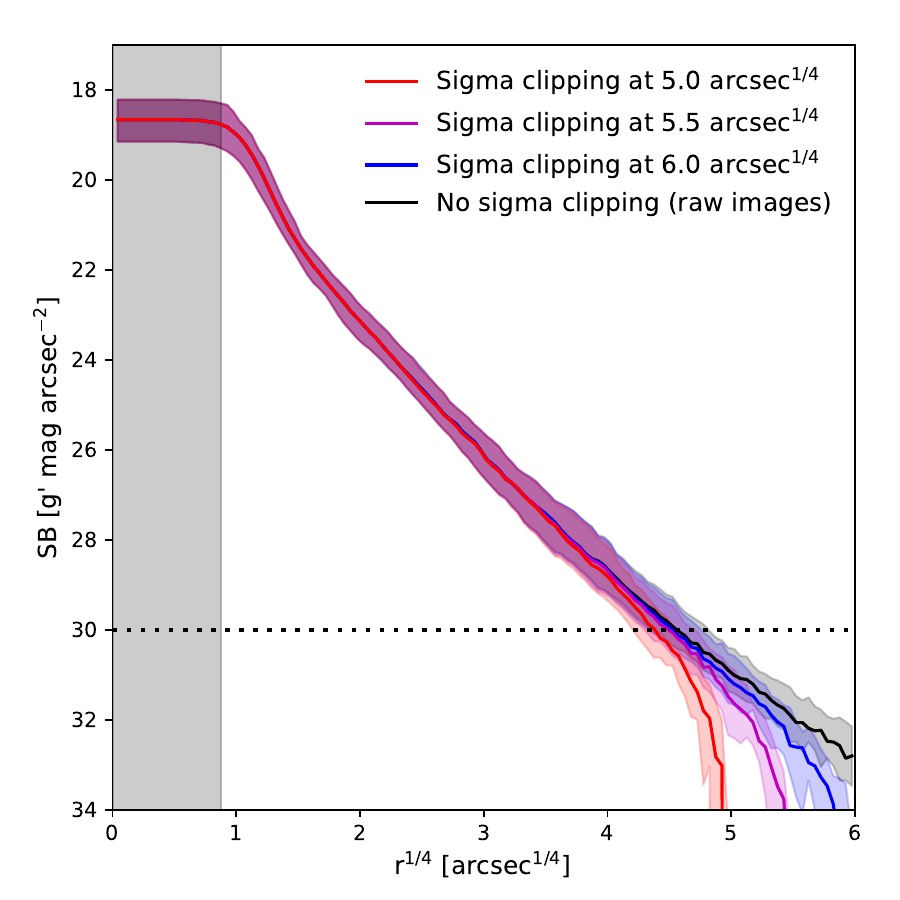}
    \includegraphics[width=9cm]{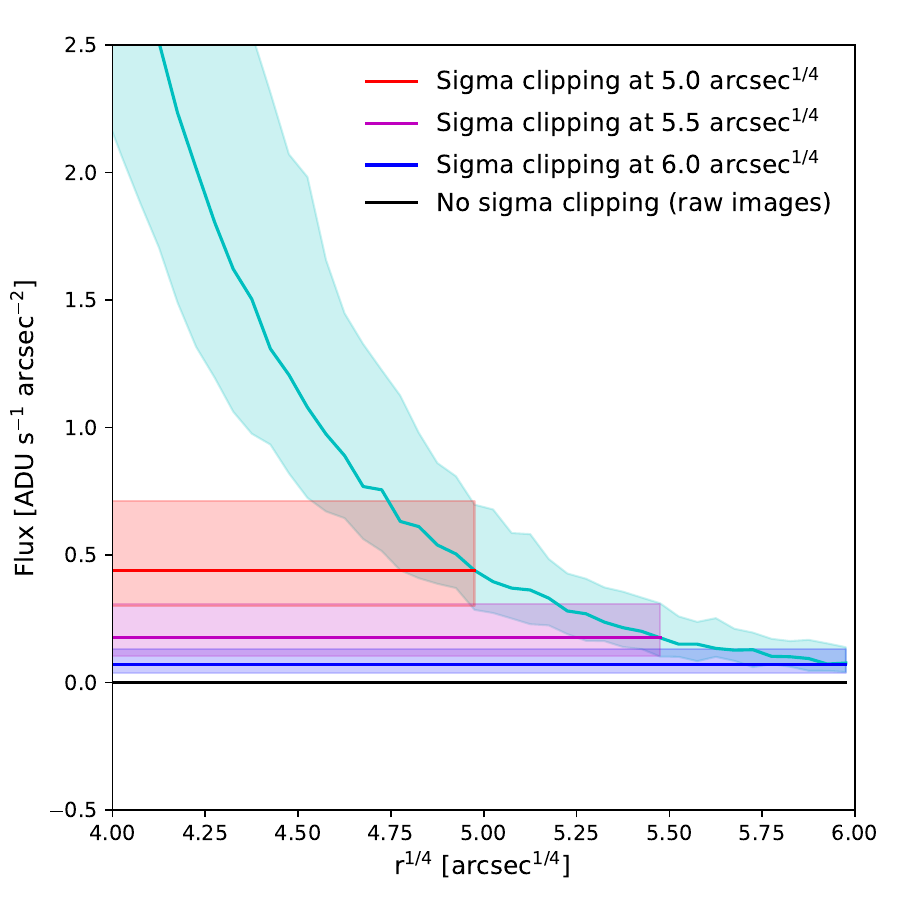}  
  }}
  \caption{Left: Median surface brightness (SB) profiles of the synthetic images, where the background levels have been obtained via sigma clipping within circular annuli at $r^{1/4} = 4.95$--$5$ arcsec$^{1/4}$ (red), $r^{1/4} = 5.45$--$5.5$ arcsec$^{1/4}$ (magenta), and $r^{1/4} = 5.95$--$6$ arcsec$^{1/4}$ (blue). The black line represents the `raw' SB profile obtained for the unmodified background (which is centred at zero, by construction). The grey shaded region to the left represents the resolution limit of the observations (0.5 times the PSF FWHM), while the horizontal dotted line indicates the limiting SB (30 $g'$ mag arcsec$^{-2}$). Right: The cyan line shows the median flux profile of the unmodified, `raw' synthetic images (which, by construction, are perfectly background subtracted), while the coloured lines indicate the background levels (determined via sigma clipping within thin annuli at various radii) used to calculate the SB profiles shown on the left-hand panel. In both panels, the coloured shaded regions indicate the 16th to 84th percentile ranges. This figure supports our choice of $r^{1/4} = 5.5$ arcsec$^{1/4}$ as an outer boundary for our main calculations, since SB values can be reliably measured down to the limiting SB within this region.}
	\label{fig:profiles_sigma_clipping}
\end{figure*}

Background subtraction is a crucial step in the analysis of low surface brightness systems, since one must be careful not to subtract the object of study itself from the image. As discussed in \cite{Kluge+2020}, even after running a customized data reduction pipeline on their WWFI observations, each of the resulting stacked images contains a residual, negative background constant that must be accounted for in post-processing. Such background constants were determined in \cite{Kluge+2020} through visual inspection of the radial flux profiles (see their fig. 10).

On the simulation side, each synthetic image has a perfectly flat background centred at zero, by construction. In order to avoid biasing our results by comparing such `perfect' background constants to the visually determined background constants from the observations, we adopt a unified, automatic procedure that can be applied to simulations and observations alike, which consists in measuring the median background level via sigma clipping within a thin annulus at a sufficiently large radius. To this end, we use the \texttt{sigma\_clipped\_stats} function of the Astropy package \citep{astropy2013, astropy2018}, clipping values that lie more than 3 standard deviations away from the median, and iterating until convergence.

Fig. \ref{fig:profiles_sigma_clipping} shows the median SB profiles of the synthetic images (left) that result from determining (and subsequently subtracting) the background constant via sigma-clipping within thin circular annuli at different radii (right). This figure shows that the `true' SB in the simulations can be recovered down to 30 $g'$ mag arcsec$^{-2}$ (the limiting SB of the observations) when the background properties are determined at sufficiently large radii. Specifically, the circular annulus between $r^{1/4} = 5.45$--$5.5$ arcsec$^{1/4}$ (magenta lines) is sufficient for our purposes. This radial range is also sufficiently large for the observations (not shown), since the 30 $g'$ mag arcsec$^{-2}$ level is typically reached at smaller radii, as will be demonstrated in Section \ref{subsec:bcg+icl_profiles}. This approach, while coarse, is straightforward and ensures a consistent characterization of the background in simulation and observations.

For convenience, we will refer to the following two types of images throughout the rest of the paper:

\begin{itemize}
    \item[\textit{Raw images:}] These refer to the original, unmodified images. We will only make use of the raw \textit{synthetic} images, which by construction are perfectly background subtracted, in order to quantify the amount of light `hidden' below the SB limit and determine the appropriate correction factors to account for it.

    \vspace{0.2cm}

    \item[\textit{Sigma-clipped images:}] These are the result of determining the median background level via sigma clipping within a thin circular annulus with radii $r^{1/4} = 5.45$--$5.5$ arcsec$^{1/4}$, as explained above, and then subtracting this constant value from the `raw' image. These will be used for our main results, allowing robust, systematic comparisons between synthetic and real images. (Note that, despite our nomenclature choice, the pixel values themselves are not sigma-clipped.)

\end{itemize}

\section{The BCG+ICL light distribution}
\label{sec:bcg+icl}

Throughout this section we characterize the BCG+ICL as a single entity, i.e. without separating the BCG from the ICL. We use the real and synthetic images to compute fitted Sérsic parameters (Section \ref{subsec:sersic_fits}), SB profiles (Section \ref{subsec:bcg+icl_profiles}), Petrosian radii (Section \ref{subsec:rpetro}), observational correction factors (Section \ref{subsec:undetected_light}) and half-light radii (Section \ref{subsec:rhalf}), carrying out apples-to-apples comparisons between the simulation and observations.

\subsection{S\'ersic fits}
\label{subsec:sersic_fits}

\begin{figure*}
  \centering
    \includegraphics[width=\linewidth]{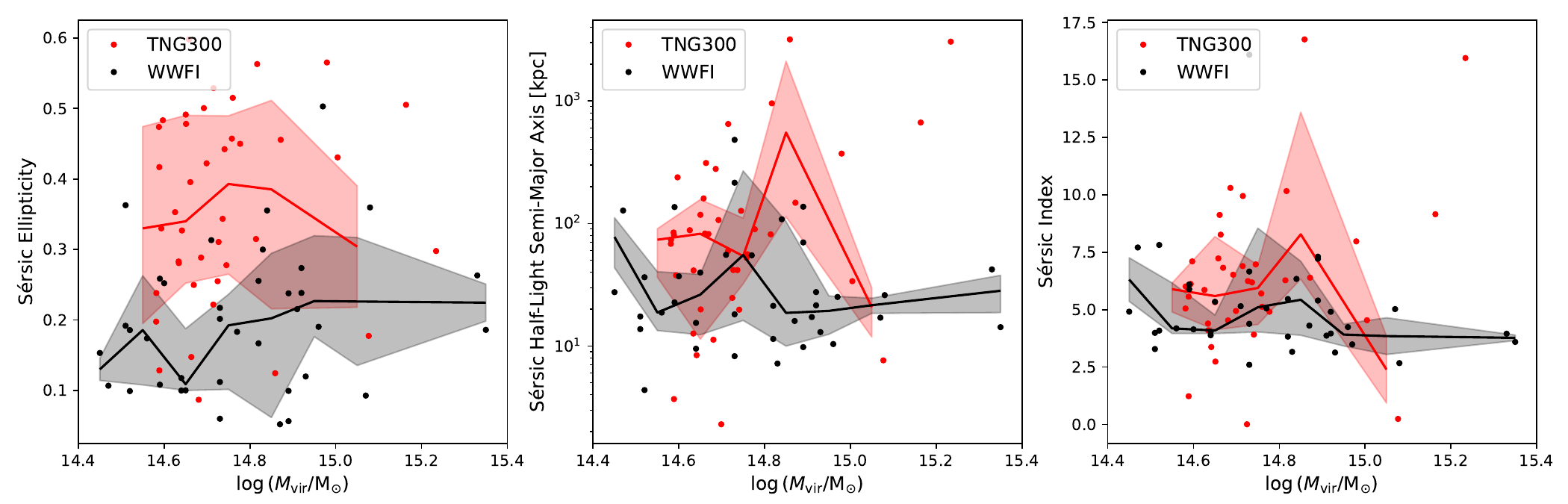}
  \caption{Ellipticities (left), sizes (middle), and Sérsic indices (right) obtained from fitting 2D Sérsic models to the synthetic (red) and real (black) BCG+ICL images, plotted against cluster mass. The solid lines and shaded regions indicate the medians and 16th-84th percentile ranges, respectively. Only observed clusters with gravitational mass estimates (38 out of 170) are shown.}
	\label{fig:morph_vs_mvir_comparison}
\end{figure*}

\begin{figure*}
  \centering
  \centerline{\hbox{
    \includegraphics[width=9cm]{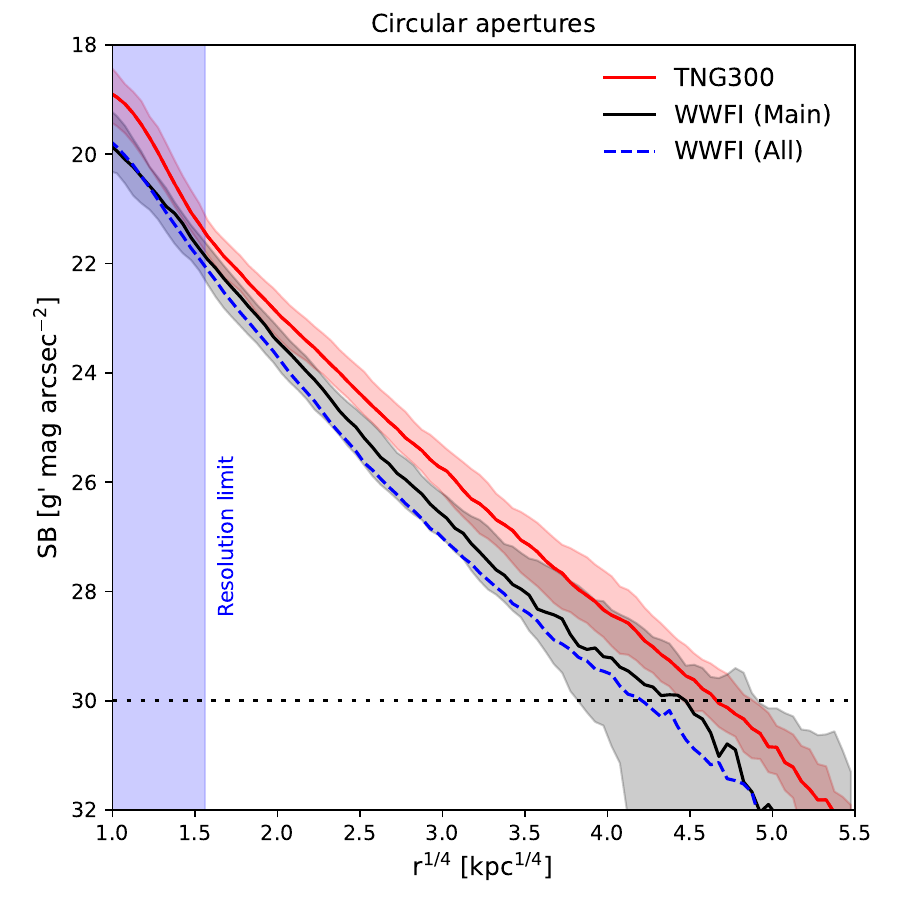}
    \includegraphics[width=9cm]{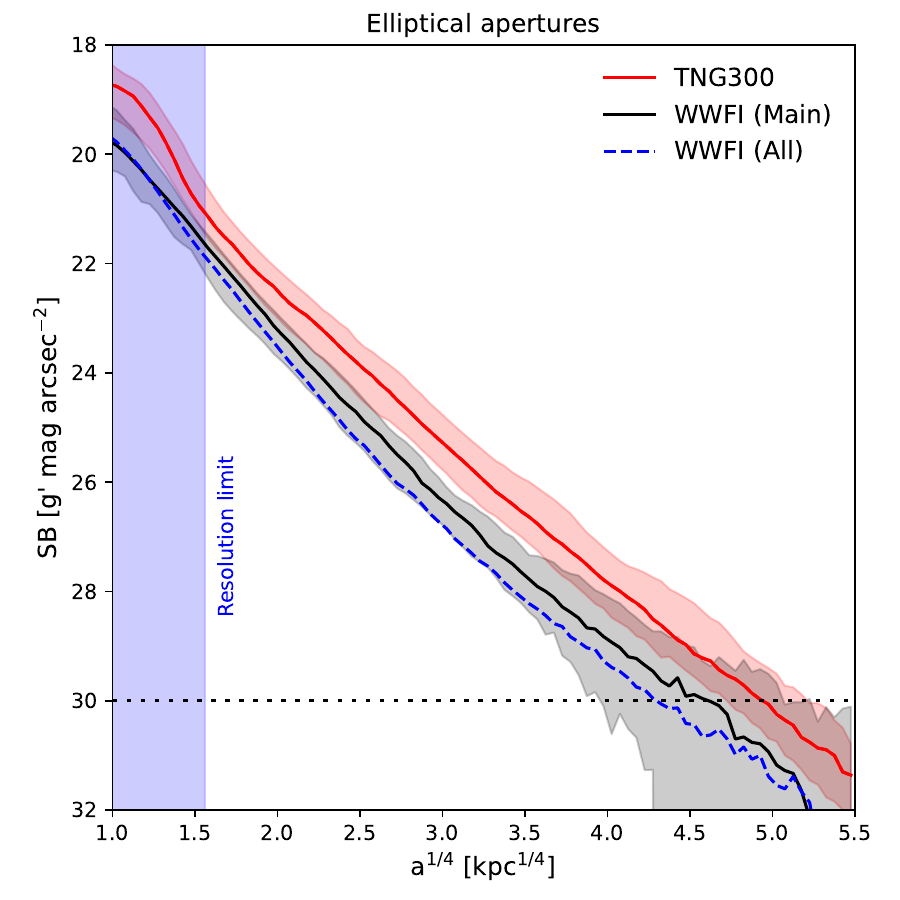}  
  }}
  \caption{Median surface brightness (SB) profiles of the simulated (red) and observed (black and blue) BCG+ICL systems, shown both for circular (left) and elliptical (right) apertures. In the case of the observations, the main (consisting of the 38 observed clusters with mass estimates) and full (including all 170 observed clusters) samples are represented by the solid black and dashed blue lines, respectively. The red and black shaded regions represent the 16th to 84th percentile ranges of the TNG300 and WWFI (main sample) profiles, respectively.  The blue shaded region to the left indicates the resolution limit of the simulation, here defined as $4 \epsilon_{\ast} \approx 5.9$ kpc. All SB values are in the observer's frame.}
	\label{fig:profiles_comparison}
\end{figure*}

\begin{figure*}
  \centering
  \centerline{\hbox{
    \includegraphics[width=9cm]{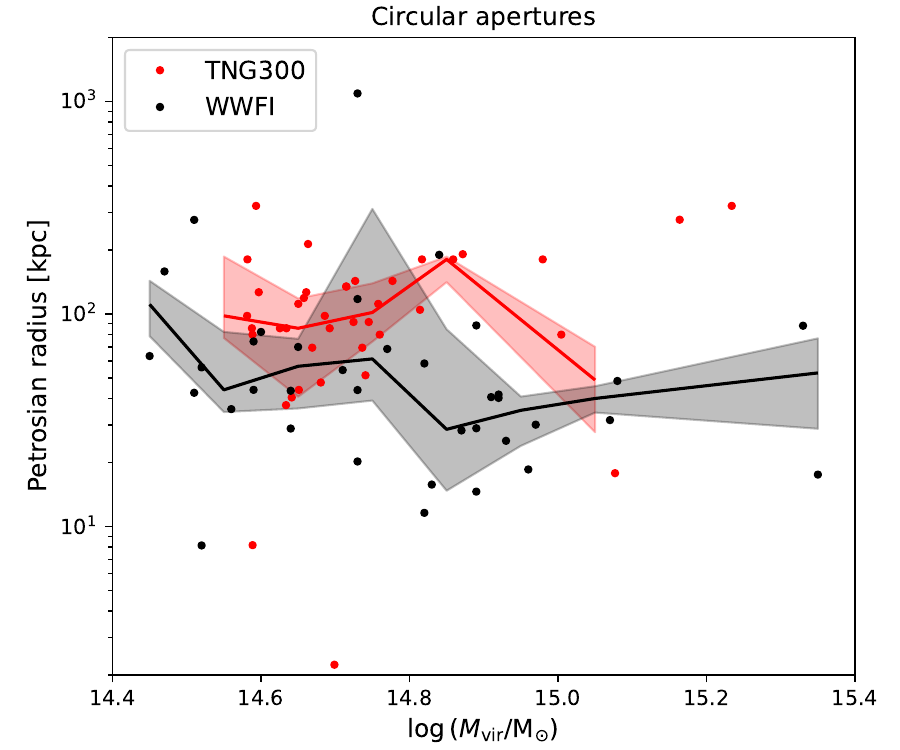}
    \includegraphics[width=9cm]{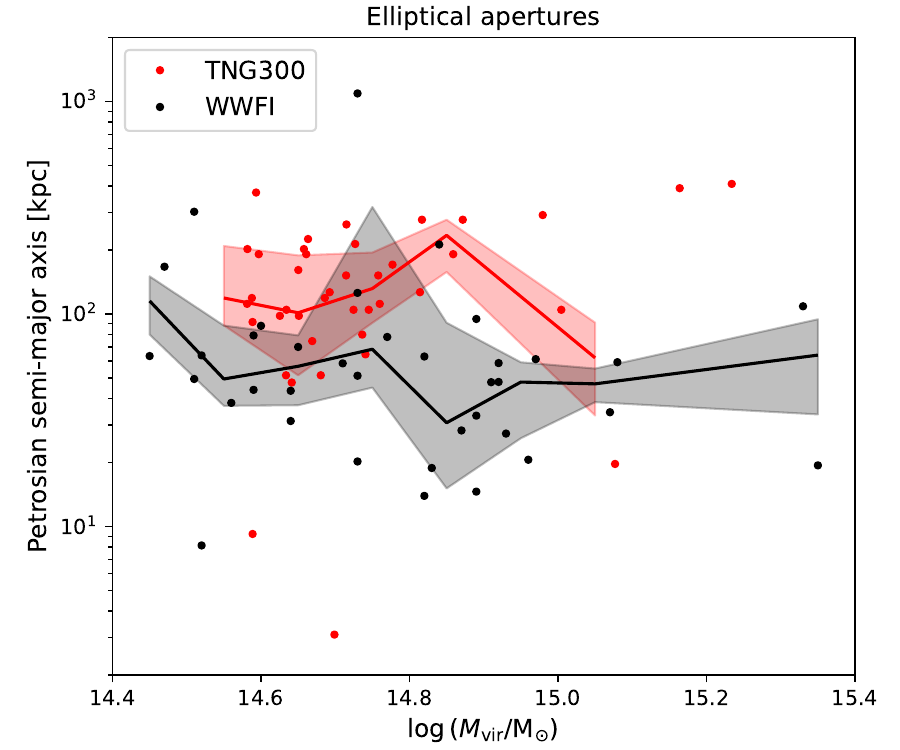}  
  }}
  \caption{Petrosian radii of the simulated (red) and observed (black) BCG+ICL systems, plotted against cluster mass. The left- and right-hand panels show measurements for circular and elliptical apertures, respectively. Solid lines indicate the medians, while shaded regions represent the 16th to 84th percentile ranges. Only observed clusters with gravitational mass estimates (38 out of 170) are shown.}
	\label{fig:rpetro}
\end{figure*}

We fit 2D Sérsic models \citep{sersic1968} to both the real and synthetic images using the \textsc{statmorph} code \citep{Rodriguez-Gomez2019}. While this tool was originally developed for images of `normal' galaxies, it can also be applied to clusters by setting an appropriately large value of the \texttt{cutout\_extent} parameter, which determines the size of the region used to carry out the computations.\footnote{More precisely, we provide as input a segmentation map where the only `source' of interest consists of a circle with radius equal to $0.1 \times 5.5^4$ arcsec $\approx 91.5$ arcsec (\textsc{statmorph} measurements are not overly sensitive to the input segmentation map), and then specify the argument \texttt{cutout\_extent} = 10, which tells \textsc{statmorph} to use the original image size ($r = 5.5^4$ arcsec). The satellite mask is provided as a separate argument.} While some of the traditional morphological measurements -- notably the minimized asymmetry parameter \citep{abraham1996, conselice2000} -- are challenging to compute in heavily masked regions such as those around BCGs in rich clusters (unless grossly simplifying assumptions are made), we find that the 2D Sérsic models returned by \textsc{statmorph} are reasonably robust, with reduced chi-squared values around 1, indicating good fits.\footnote{Note that the fitted 2D Sérsic models have a fixed ellipticity and position angle at all radii.}

Fig. \ref{fig:morph_vs_mvir_comparison} shows the fitted parameters of the 2D Sérsic models as a function of cluster mass, comparing the TNG300 simulated clusters (red) with the WWFI observed clusters (black). This figure shows that the simulated clusters tend to have higher ellipticities, with median values and scatter (here defined as half of the 16th to 84th percentile range) of $0.35 \pm 0.14$, while the observed systems have lower values, of $0.19 \pm 0.10$. Similarly, the simulated clusters have larger (Sérsic-inferred) sizes than the observations, with $\log_{10}\left(r_{\rm e} \, [{\rm kpc}]\right) = 1.9 \pm 0.6$ versus $\log_{10}\left(r_{\rm e} \, [{\rm kpc}]\right) = 1.3 \pm 0.4$. These size measurements should be interpreted with caution, since they are model-dependent (non-parametric measurements of BCG+ICL sizes will be explored in Sections \ref{subsec:rpetro} and \ref{subsec:rhalf}). The Sérsic indices are also higher in the simulation than in the observations, with values of $6.1 \pm 2.4$ and $4.3 \pm 1.4$, respectively.\footnote{Unless otherwise noted, the reported statistics (median and scatter) refer to the 38 observed clusters with gravitational mass estimates. The statistics would be very similar for the full sample of 170 objects.} Importantly, we do not observe a strong dependence on cluster mass for any of the fitted parameters, at least within the mass range considered.

\subsection{Surface brightness profiles}
\label{subsec:bcg+icl_profiles}

Fig. \ref{fig:profiles_comparison} shows the median SB profiles (as a function of $r^{1/4}$, following the de Vaucouleurs law) obtained for the TNG300 simulation (red) and WWFI observations (black and blue), computing SB values in the observer's frame using circular (left) and elliptical (right) apertures. For the latter, we use the ellipticity and orientation obtained from the fitted Sérsic models (Section \ref{subsec:sersic_fits}). While it is known that the ellipticity of the BCG+ICLs is radius-dependent, with isophotes becoming more elliptical at larger radii \citep[e.g.][]{Kluge+2020}, here we assume a fixed ellipticity value for the sake of simplicity, since we are mainly interested in comparing theory versus observations rather than studying in detail how isophotal properties change with radius. 

As seen from Fig. \ref{fig:profiles_comparison}, there is a clear offset between the simulated and observed SB profiles, with the former appearing brighter than the latter. The discrepancy is somewhat larger for the full sample of 170 observed clusters (dashed blue lines) than for the main sample of 38 observed clusters with gravitational mass estimates (solid black lines). Focusing on the latter, the magnitude of the difference between simulation and observations is minimal around $r^{1/4} \approx 1.6$ kpc$^{1/4}$, which coincides with the spatial resolution limit of the simulation, here defined as four times the gravitational softening length of stars and DM, $4 \epsilon_{\ast} \approx 5.9$ kpc. Around this point, the simulated profiles are brighter than the observed ones by around 0.4 $g'$ mag arcsec$^{-2}$ and 0.6 $g'$ mag arcsec$^{-2}$ for circular and elliptical apertures, respectively. This difference becomes larger with radius, since the TNG300 profiles are somewhat shallower than the observed ones at $r^{1/4} \approx 2$--$3$ kpc$^{1/4}$, although the slopes become roughly parallel at larger radii, at $r^{1/4} \gtrsim 3$ kpc$^{1/4}$ (or $r \gtrsim 100$ kpc). In this regime, the difference between the median SB profiles becomes approximately constant at 0.9 $g'$ mag arcsec$^{-2}$ and 1.1 $g'$ mag arcsec$^{-2}$ for circular and elliptical apertures, respectively. The scatter, on the other hand, is similar between the simulation and observations, except at sufficiently large radii ($r^{1/4} \gtrsim 4$ kpc$^{1/4}$) where the SB limit starts to significantly impact the observations.

The fact that the simulated SB profiles are systematically brighter than the observed ones does not necessarily have an impact on \textit{relative} measurements of the ICL, such as the ICL fractions that will be the focus of Section \ref{sec:icl_fraction}. In Section \ref{sec:summary_and_discussion} we will discuss the amplitude differences between the observed and predicted SB profiles.

\subsection{Petrosian radii}
\label{subsec:rpetro}

\begin{figure*}
  \centering
  \centerline{\hbox{
    \includegraphics[width=9cm]{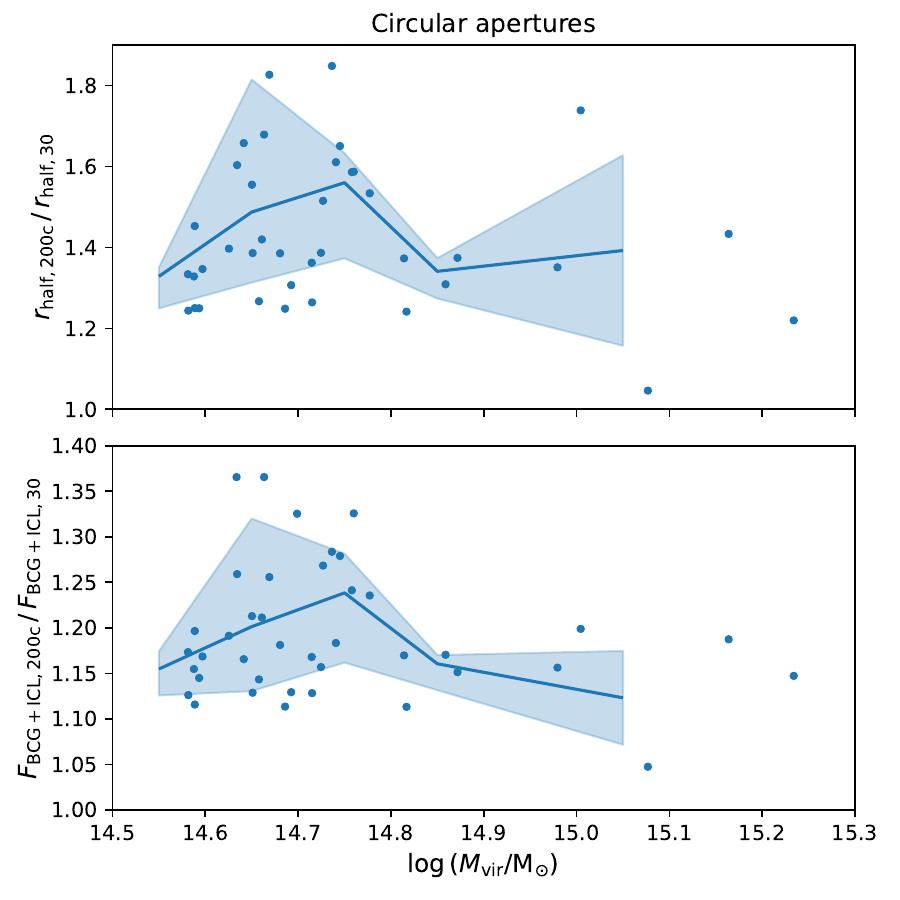}
    \includegraphics[width=9cm]{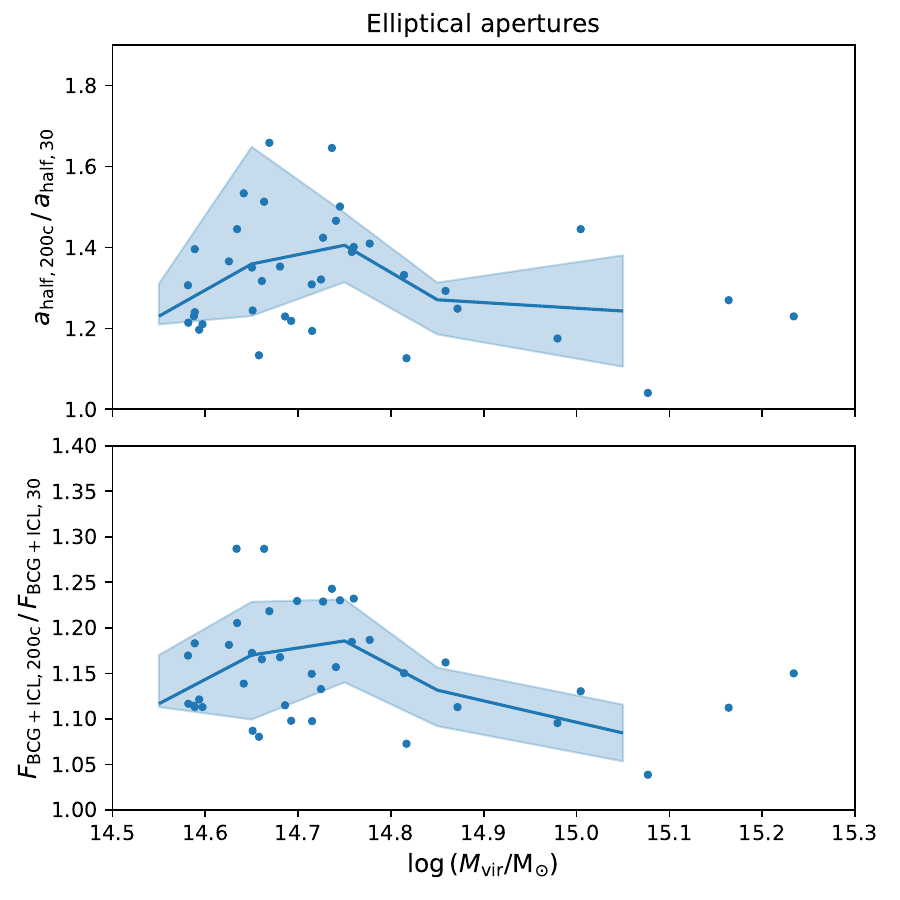}  
  }}
  \caption{Correction factors for half-light radii (top) and BCG+ICL fluxes (bottom), i.e. the multiplicative factors required to convert a `SB-limited' measurement (where all quantities are measured within the observational limit of 30 $g'$ mag arcsec$^{-2}$) into a `full' measurement (where all fluxes are measured out to $r_{\rm 200, crit}$, which is only possible to do for the `raw' synthetic images). Top left: Ratio between the `full' half-light radius and the `SB-limited' half-light radius, plotted against cluster mass. Bottom left: Ratio between the `full' BCG+ICL flux and the `SB-limited' BCG+ICL flux, plotted against cluster mass. Top right: Same as top left, but for elliptical apertures. Bottom right: Same as bottom left, but for elliptical apertures. In each panel, the solid line indicates the median, while the shaded region indicates the 16th to 84th percentile range. All measurements were obtained for the `raw' synthetic images (see Section \ref{subsec:background_subtraction}).}
	\label{fig:correction_vs_mvir}
\end{figure*}

The Petrosian radius \citep{Petrosian1976} is defined as the radius where the mean SB is 0.2 times the mean SB within that radius. It is a very useful size measurement because it is independent of the light distribution beyond that radius, and is also largely insensitve to the observation depth.

Fig. \ref{fig:rpetro} shows the Petrosian radii of the simulated (red) and observed (black) BCG+ICL systems as a function of cluster mass, calculated for circular (left) and elliptical (right) apertures. Interestingly, the Petrosian radii are not strongly sensitive to cluster mass, at least within the explored mass range. The simulated systems are substantially larger than the observed ones, with median values of 98 kpc (123 kpc) for TNG300 and 43 kpc (50 kpc) for WWFI using circular (elliptical) apertures. This represents a size ratio of 2.3 (2.5) for the circular (elliptical) measurements. The scatter (again defined as half of the 16th to 84th percentile range) is approximately 0.3 dex for both the simulation and observations. As with the SB profiles, the differences between simulated and observed BCG+ICL sizes do not necessarily have an impact on the ICL fractions that will be discussed in Section \ref{sec:icl_fraction}. The differences between the observed and simulated BCG+ICL sizes will also be discussed in Section \ref{sec:summary_and_discussion}.

\subsection{Undetected light below the SB limit}
\label{subsec:undetected_light}

Before calculating the half-light radius (Section \ref{subsec:rhalf}), it is important to estimate the amount of undetected light below the SB limit, since this affects the measurement of the \textit{total} light of the BCG+ICL, which is a crucial part of computing the half-light radius.

The SB limit of the WWFI observations is 30 $g'$ mag arcsec$^{-2}$, which means that fluxes cannot be reliably measured beyond this point. \cite{Kluge+2020} addressed this issue by extrapolating their fitted Sérsic profiles to infinity and thus estimating the total BCG+ICL luminosity. Here, on the other hand, we compute effective correction factors by comparing two different types of measurements:

\begin{itemize}
    \item[\textit{SB-limited measurements:}] These are the result of integrating fluxes out to the SB limit of 30 $g'$ mag arcsec$^{-2}$, which is provisionally assumed to contain all the light of the BCG+ICL. These measurements can be carried out on both the synthetic and real images, and are subsequently `extrapolated' to $r_{\rm 200, crit}$ using the correction factors presented later in this section.

    \vspace{0.2cm}

    \item[\textit{Full measurements:}] These are the result of integrating fluxes out to $r_{\rm 200, crit}$, which we consider to be a physically motivated outer limit for the BCG+ICL. These measurements are only meaningful when carried out on the `raw' synthetic images (see Section \ref{subsec:background_subtraction}), as they have perfectly flat backgrounds centred at zero (although a realistic amount of background noise is included, it is homogeneously distributed, so that the fluctuations cancel out when calculating integrated fluxes).

\end{itemize}

\begin{figure*}
  \centering
  \centerline{\hbox{
    \includegraphics[width=9cm]{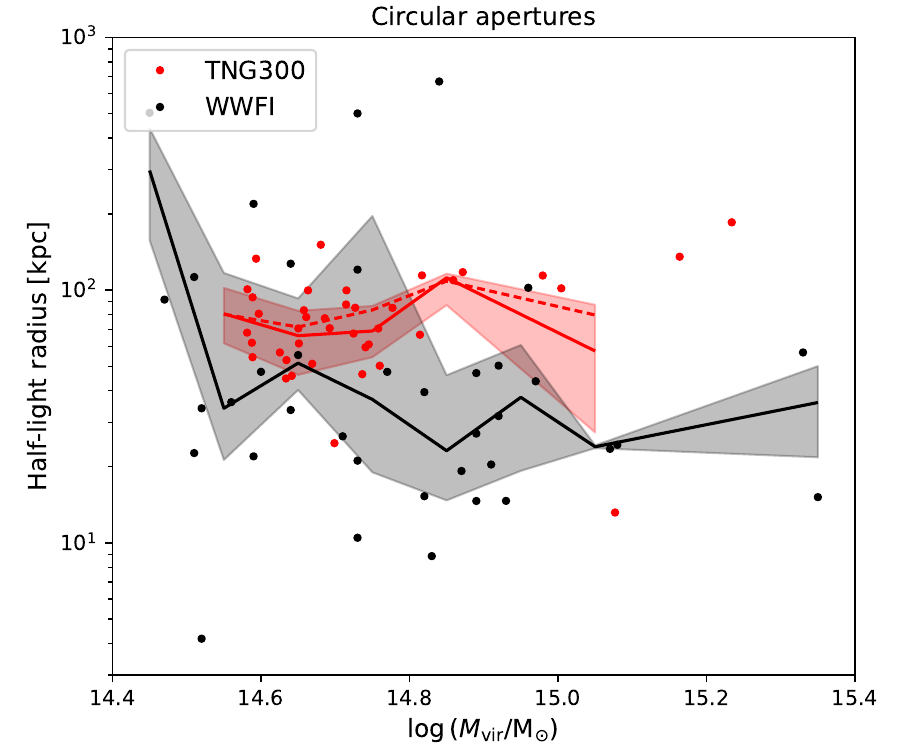}
    \includegraphics[width=9cm]{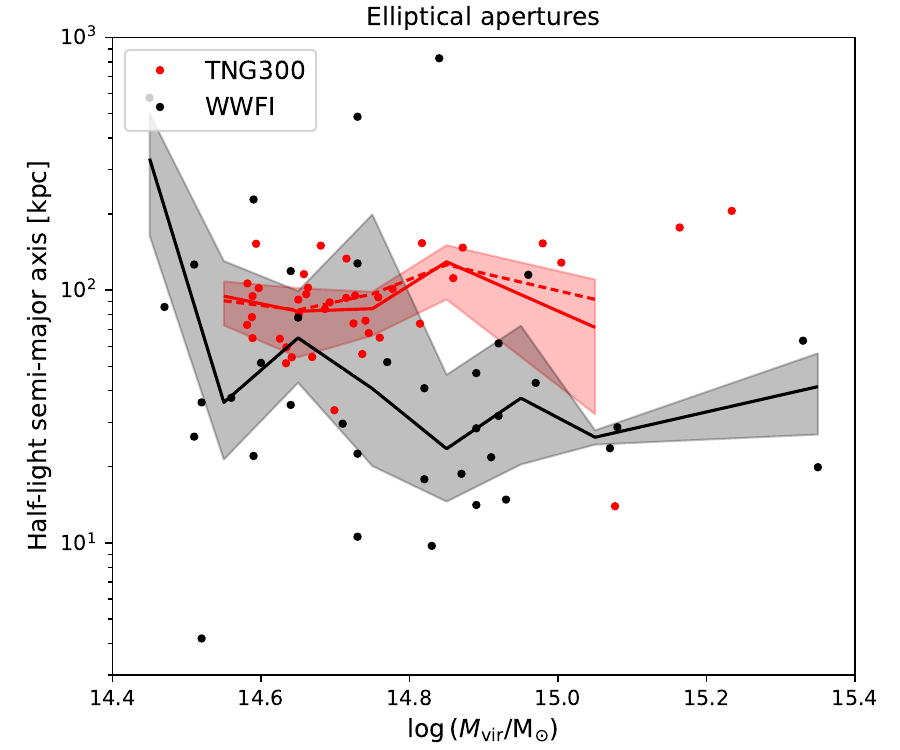}  
  }}
  \caption{Half-light radii of the simulated (red) and observed (black) BCG+ICL systems, plotted against cluster mass. The left- and right-hand panels correspond to circular and elliptical apertures, respectively. The dots show the corrected, `SB-limited' measurements (see Section \ref{subsec:undetected_light}) obtained for the `sigma-clipped' images (see Section \ref{subsec:background_subtraction}), while the solid lines and shaded regions show the corresponding medians and 16th--84th percentile ranges, respectively. The dashed red line shows the median result obtained from performing `full' measurements (see Section \ref{subsec:undetected_light}) on the `raw' synthetic images (see Section \ref{subsec:background_subtraction}). In other words, the solid and dashed red lines represent the approximate (observationally feasible) and true (simulation-only) `full' measurements, respectively. Only observed clusters with gravitational mass estimates (38 out of 170) are shown.}
	\label{fig:rhalf}
\end{figure*}

Fig. \ref{fig:correction_vs_mvir} shows the correction factors obtained from calculating the ratio between the quantity of interest -- half-light radius (top) or total BCG+ICL flux (bottom) -- measured out to $r_{\rm 200, crit}$ (`full' measurement) dividided by the same quantity measured out to the 30 $g'$ mag arcsec$^{-2}$ limit (`SB-limited' measurement) in the `raw' synthetic images, plotted against cluster mass. The left-hand and right-hand panels correspond to circular and elliptical apertures, respectively. Note that none of the correction factors is strongly dependent on cluster mass.

The median values and scatter -- here defined as half of the 16th to 84th percentile range -- of the correction factors for \textit{circular} apertures are the following:
\begin{equation}
  C_{\rm rhalf, circ} = r_{\rm half, 200c} \, / \, r_{\rm half, 30} = 1.39 \pm 0.20,
  \label{eq:correction_rhalf_circ}
\end{equation}
\vspace{-0.6cm}
\begin{equation}
  C_{\rm flux, circ} = F_{\rm BCG+ICL, 200c} \, / \, F_{\rm BCG+ICL, 30} = 1.17 \pm 0.07.
  \label{eq:correction_flux_circ}
\end{equation}
Similarly, for \textit{elliptical} apertures:
\begin{equation}
  C_{\rm rhalf, ellip} = a_{\rm half, 200c} \, / \, a_{\rm half, 30} = 1.32 \pm 0.14,
  \label{eq:correction_rhalf_ellip}
\end{equation}
\vspace{-0.6cm}
\begin{equation}
  C_{\rm flux, ellip} = F_{\rm BCG+ICL, 200c} \, / \, F_{\rm BCG+ICL, 30} = 1.15 \pm 0.06,
  \label{eq:correction_flux_ellip}
\end{equation}

The latter are somewhat larger than the median correction factors obtained by \cite{Kluge+2020} from extrapolating Sérsic fits to the observed systems: $1.20 \pm 0.15$ for half-light radii and $1.09 \pm 0.06$ for BCG+ICL fluxes (see their equations 33 and 35), although the scatter obtained by \cite{Kluge+2020} is almost exactly the same as the one we obtain for the synthetic images. At face value, the differences in the median correction factors suggest that the far-out SB profiles in the simulation become slightly shallower than the extrapolated Sérsic functions in the observations. This could be either due to intrinsically different profiles or due to overestimations of the observational background.

\begin{figure*}
  \centering
  \centerline{\hbox{
    \includegraphics[width=9cm]{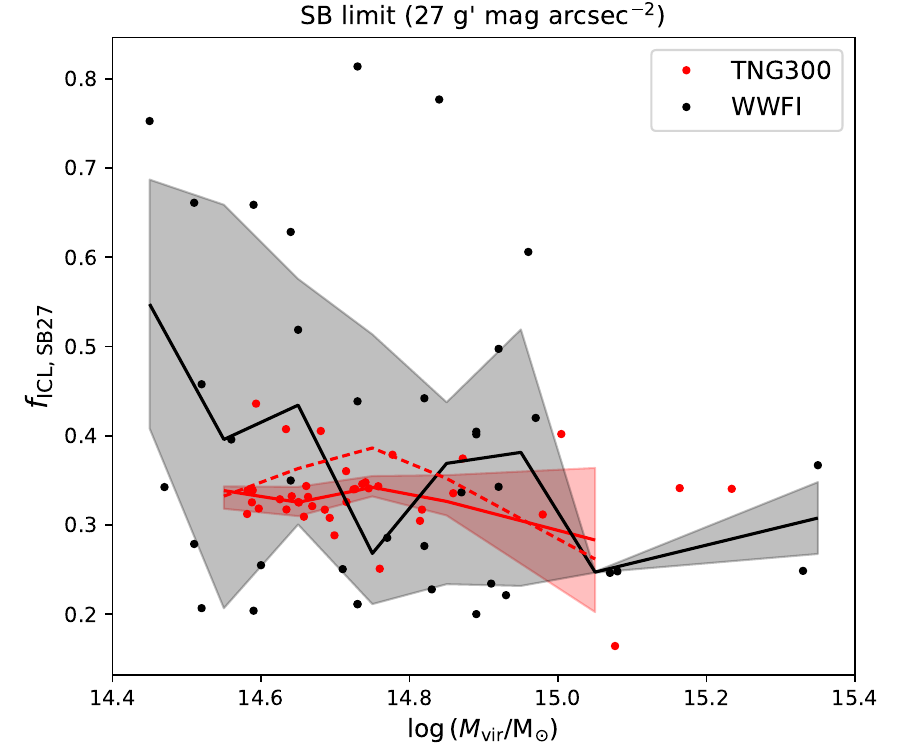}
    \includegraphics[width=9cm]{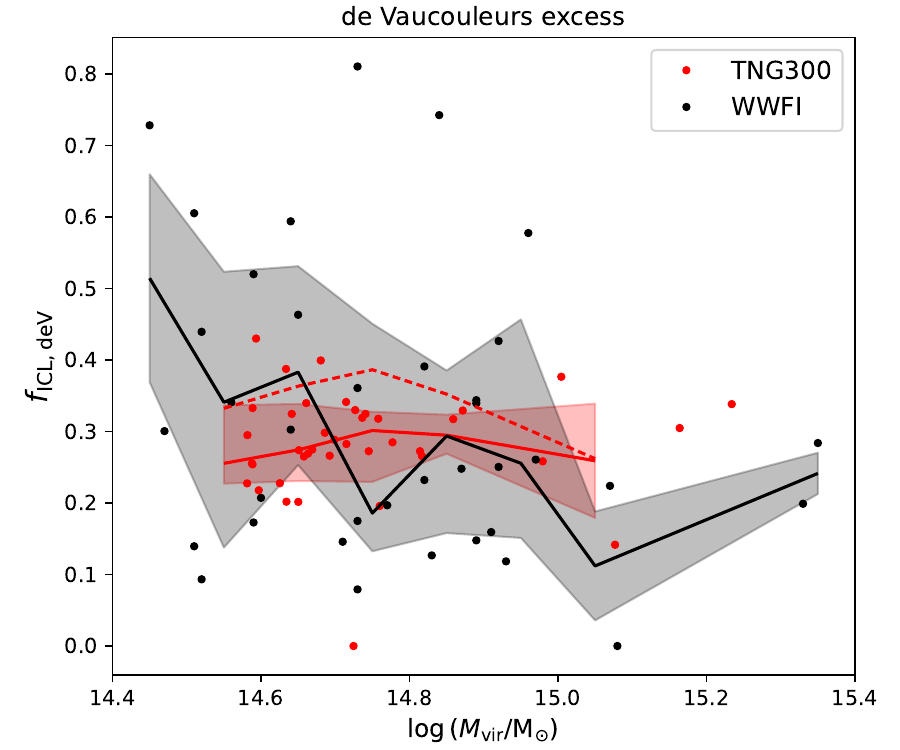}  
  }}
	\caption{The ICL fraction that results from defining the ICL as the light with SB dimmer than 27 $g'$ mag arcsec$^{-2}$ (left) and as the excess of light with respect to an outwardly extrapolated 2D de Vaucouleurs profile fitted to the inner region brighter than 27 $g'$ mag arcsec$^{-2}$ (right), plotted against cluster mass. Red and black dots represent the simulation and observations, respectively, while the solid lines and shaded regions indicate the corresponding medians and 16th--84th percentile ranges. As before, the main measurements (dots, solid lines, and shaded regions) represent the corrected, `SB-limited' measurements (see Section \ref{subsec:undetected_light}) obtained for the `sigma-clipped' images (see Section \ref{subsec:background_subtraction}), while the dashed red line shows the median $f_{\rm ICL}$ obtained by performing `full' measurements on the `raw' synthetic images. Only observed clusters with gravitational mass estimates (38 out of 170) are shown.}
	\label{fig:f_icl_sblim_27_and_dev}
\end{figure*}

From this point onward, in order to ensure a fair comparison between theory and observations, we apply the median correction factors from equations (\ref{eq:correction_rhalf_circ})--(\ref{eq:correction_flux_ellip}) to the `SB-limited' measurements (out to 30 $g'$ mag arcsec$^{-2}$) for \textit{both} the synthetic and real `sigma-clipped' images (see Section \ref{subsec:background_subtraction}). These corrected, `SB-limited' measurements (presented for both simulation and observations) can be thought of as an observationally feasible approximation of the true `full' measurements (only achievable in the simulation). However, in order to assess possible biases in the correction factors from equations (\ref{eq:correction_rhalf_circ})--(\ref{eq:correction_flux_ellip}) due to individual variation between clusters, we also show the true `full' measurements (i.e. integrating the fluxes out to $r_{\rm 200, crit}$) obtained directly from the `raw' synthetic images (see Section \ref{subsec:background_subtraction}) as \textit{dashed} lines in most of the upcoming plots (Figs \ref{fig:rhalf}--\ref{fig:f_icl_fixed_kpc}).

\subsection{Half-light radii}
\label{subsec:rhalf}

Unlike for the Petrosian radius (Section \ref{subsec:rpetro}), computing the (non-parametric) half-light radius ($r_{\rm half}$) has the disadvantage of needing to define an outer radius that is assumed to contain the `total' light of the object (unless a parametric model is extrapolated to infinity, which also involves several assumptions). How to define such an outer radius is often not a major point of discussion for `normal' galaxies, which have steeply declining surface brightness profiles at large radii (and therefore a reasonably well defined `total' amount of light for sufficiently large apertures), but becomes crucial for BCG+ICLs, which exhibit very extended surface brightness profiles, and therefore contain a non-negligible amount of light `hidden' beyond the reach of even the deepest cluster observations currently available.

In order to adhere to observational SB limits, we calculate `SB-limited' half-light radii (see Section \ref{subsec:undetected_light}) for \textit{both} the real and synthetic images, and then apply the correction factors from equations (\ref{eq:correction_rhalf_circ}) and (\ref{eq:correction_rhalf_ellip}) in order to estimate the `full' half-light radii. These corrected, `SB-limited' measurements are shown in Fig. \ref{fig:rhalf} as dots (along with their median and scatter as solid lines and shaded regions), where red corresponds to the simulation and black to the observations. The left- and right-hand panels show measurements for circular and elliptical apertures, respectively. For reference, the dashed red lines show the `full' measurements (see Section \ref{subsec:undetected_light}) obtained directly for the `raw' synthetic images (see Section \ref{subsec:background_subtraction}). Note that the solid and dashed red lines are very close to each other, which means that the median correction factors introduced in Section \ref{subsec:undetected_light} are reasonably accurate.

For the range of masses of our simulated clusters, there is no significant dependence of the half-light radius on $M_{\rm vir}$.\footnote{Although there seems to be some anticorrelation between the half-light radius (or semi-major axis) and the cluster mass, we find that it is not statistically significant. Specifically, we obtain a Spearman's rank correlation coefficient of $-0.147$ with a \textit{p}-value of $0.378$, indicating that there is a nearly 40 per cent probability that a correlation of this magnitude (or larger) occurs by chance in an uncorrelated dataset. This \textit{p}-value is larger than the threshold of 0.05 typically used to assess statistical significance.} The simulated clusters are again larger than the observed ones, with median half-light radii of 74 kpc (92 kpc) for TNG300 and 34 kpc (36 kpc) for WWFI using circular (elliptical) apertures. This represents a median size ratio of 2.2 (2.6) for the circular (elliptical) measurements, which is consistent with the median size ratios obtained for the Petrosian radii (Section \ref{subsec:rpetro}). The scatter is approximately 0.2 dex for the simulation and 0.4 dex for the observations. The latter is larger due to observational uncertainties in determinations of the total BCG+ICL flux, as will become clear in later sections.

\section{The ICL fraction}
\label{sec:icl_fraction}

The ICL fraction is defined as the ratio between the flux from the ICL and the flux from the BCG+ICL:
\begin{equation}
    f_{\rm ICL} = \frac{F_{\rm ICL}}{F_{\rm BCG + ICL}}.
    \label{eq:f_icl}
\end{equation}
As discussed in Section \ref{subsec:undetected_light}, BCG+ICL fluxes cannot be reliably measured beyond the SB limit (30 $g'$ mag arcsec$^{-2}$) in the observations. However, they can be measured out to the SB limit and then `extrapolated' out to $r_{\rm 200, crit}$ using the correction factors given by equations (\ref{eq:correction_flux_circ}) and (\ref{eq:correction_flux_ellip}). Taking this into account, equation (\ref{eq:f_icl}) becomes
\begin{equation}
    f_{\rm ICL} = 1 - \frac{F_{\rm BCG}}{F_{\rm BCG + ICL}} = 1 - \frac{F_{\rm BCG}}{ C_{\rm flux} F_{\rm BCG + ICL, 30}},
    \label{eq:f_icl_corrected}
\end{equation}
where it has been assumed that the BCG flux ($F_{\rm BCG}$) is unchanged by the choice of outer radius.\footnote{Strictly speaking, this is not true for the de Vaucouleurs method discussed in Section \ref{subsec:f_icl_dev}, since there is spatial overlap between the BCG and ICL components and therefore a nonzero BCG component beyond 30 $g'$ mag arcsec$^{-2}$. However, we find that the `SB-limited' $F_{\rm BCG, 30}$ and the `full' $F_{\rm BCG}$ measurements differ by less than 1 per cent, resulting in a negligible BCG flux correction.}

In general, the corrected, `SB-limited' measurements of $f_{\rm ICL}$, as given by equation (\ref{eq:f_icl_corrected}), are carried out on the `sigma-clipped' images (see Section \ref{subsec:background_subtraction}) and are presented throughout this section as \textit{solid} lines for both simulations and observations. For reference, we will also show the true, `full' measurements of $f_{\rm ICL}$ obtained by directly applying equation (\ref{eq:f_icl}) on the `raw' synthetic images, which will be displayed as \textit{dashed} lines in the upcoming plots.

\subsection{Surface brightness cut}
\label{subsec:f_icl_sb27}

\begin{figure*}
  \centering
  \centerline{\hbox{
    \includegraphics[width=9cm]{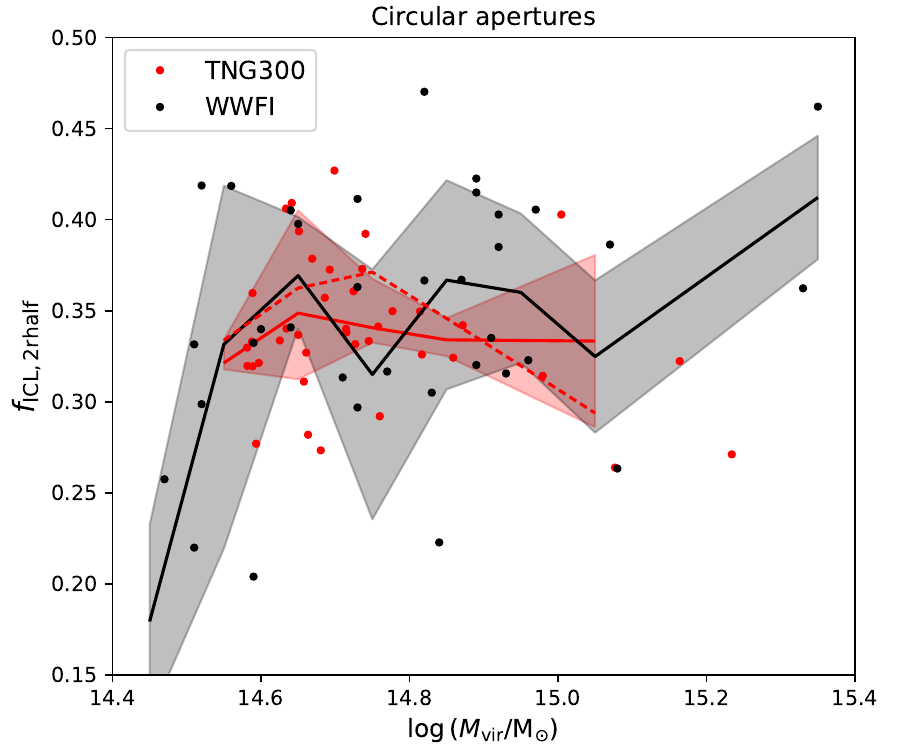}
    \includegraphics[width=9cm]{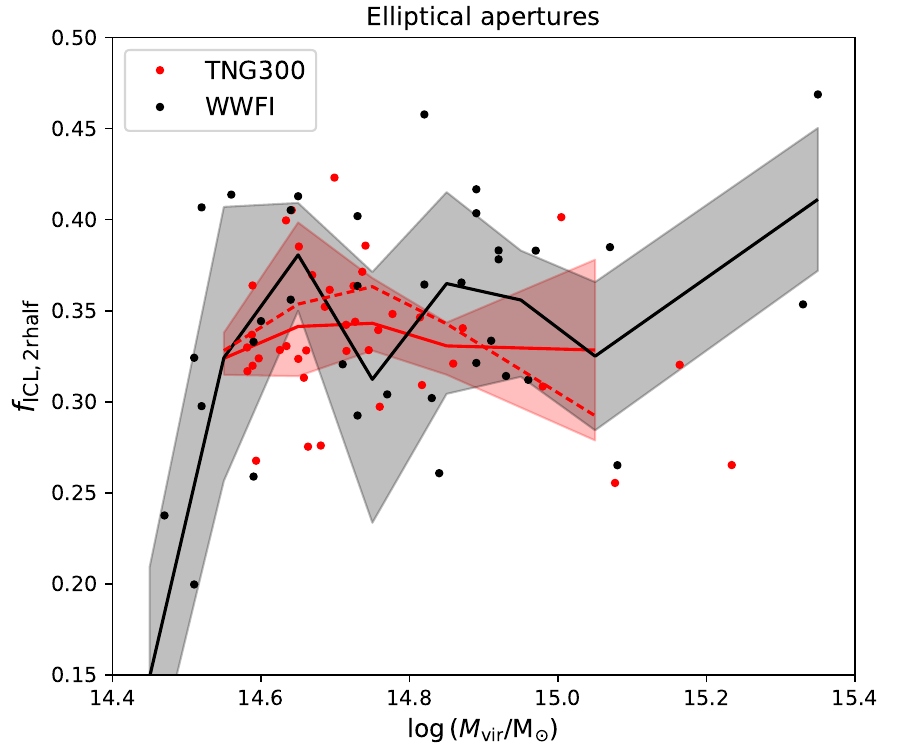}  
  }}
  \caption{The ICL fraction for TNG300 (red) and WWFI (black), where the ICL is defined as the light beyond twice the half-light radius, plotted against cluster mass. The left- and right-hand panels show results for circular and elliptical apertures, respectively. Dots represent the corrected, `SB-limited' measurements carried out for the `sigma-clipped' images, while the solid lines and shaded regions indicate the corresponding medians and percentile ranges. The dashed lines show the median ICL fractions obtained by performing `full' measurements directly on the `raw' synthetic images. Only observed clusters with gravitational mass estimates (38 out of 170) are shown.}
	\label{fig:f_icl_2rhalf}
\end{figure*}

One straightforward way of defining the ICL fraction of a BCG+ICL system is to define a SB threshold such that all light dimmer than it is assigned to the ICL. Here we follow \cite{kluge2021photometric} and adopt a threshold of 27 $g'$ mag arcsec$^{-2}$. While the main calculation is fundamentally the same, we highlight two methodological differences between our implementation and that of \cite{kluge2021photometric}:

\begin{itemize}
    \item[(i)] We apply the SB threshold in the \textit{observer's frame}, which effectively `pushes' the 27 $g'$ mag arcsec$^{-2}$ contour slightly inward compared to applying the threshold in the rest frame. However, since the 30 $g'$ mag arcsec$^{-2}$ level (by definition, the outer limit in the `SB-limited' calculations) is also `pushed' slightly inward, the net effect on the ICL fraction partly cancels out. Similarly, we do not expect a large effect on the scatter of the ICL fractions from measuring fluxes in the observer's frame, since, as discussed in Section \ref{subsec:observations}, the observed sources are located within a relatively narrow redshift range, $z \approx 0.06 \pm 0.02$, which does not introduce a large amount of scatter in the SB values.

    \vspace{0.2cm}

    \item[(ii)] More importantly, while \cite{kluge2021photometric} computed SB profiles using the \texttt{ellfitn} code \citep{Bender&Moellenhoff1987}  and then applied the SB cut on the resulting 1D profiles, here we apply the SB cut directly on the 2D images. In order to obtain a reasonably clean contour at 27 $g'$ mag arcsec$^{-2}$ in an otherwise noisy image, we first detect sources using \texttt{photutils} \citep{bradley2021_photutils} with a threshold equivalent to 27 $g'$ mag arcsec$^{-2}$, then select the largest continuous segment and compute its `effective' radius as $r_{27} = \sqrt{A_{27} / \pi}$, where $A_{27}$ is the area of the segment, and finally smooth the shape of the segment by convolving with a uniform filter with size equal to $0.2 \, r_{27}$. This `regularized' contour is then superposed on the original image in order to compute the fluxes from the BCG and the ICL (excluding satellites), allowing us to determine the ICL fraction. The outer boundary is an appropriately orientated ellipse at the SB limit of 30 $g'$ mag arcsec$^{-2}$ (see Sections \ref{subsec:sersic_fits}--\ref{subsec:bcg+icl_profiles}). The flux correction for elliptical apertures (see Section \ref{subsec:undetected_light}) is applied to equation (\ref{eq:f_icl_corrected}).
\end{itemize}

The left-hand panel of Fig. \ref{fig:f_icl_sblim_27_and_dev} shows the ICL fractions obtained from applying the method described above to the synthetic (red) and real (black) images, plotted as a function of cluster mass. The median and scatter (defined as half of the 16th--84th percentile range) of the two samples are $f_{\rm ICL, SB27} = 0.33 \pm 0.02$ for the simulation and $f_{\rm ICL, SB27} = 0.34 \pm 0.19$ for the observations. The latter turns out to be exactly equal to the result reported by \cite{kluge2021photometric} for the same observations, despite the methodological differences outlined above.

It might seem surprising that the \textit{median} $f_{\rm ICL, SB27}$ shows very good agreement between TNG300 and WWFI, despite the systematic offset between the simulated and observed SB profiles seen in Fig. \ref{fig:profiles_comparison}. The reason for this is that, while the 27 $g'$ mag arcsec$^{-2}$ contours (which represent the BCG/ICL boundary) are more extended in the simulation than in the observations, the same applies to the 30 $g'$ mag arcsec$^{-2}$ contours (which represent the total BCG+ICL extent). This results in both the ICL and BCG+ICL fluxes being larger in TNG300 than in the observations, but these systematic differences largely cancel out when computing the flux ratio.

On the other hand, the \textit{scatter} in $f_{\rm ICL, SB27}$ shows a large discrepancy between TNG300 and WWFI, with observational measurements exhibiting more variation by about an order of magnitude. While this discrepancy could potentially reflect physical processes not fully captured in the simulation, we believe it can be mainly attributed to differences in the amount of integrated BCG+ICL flux within the SB limit. Looking again at the SB profiles from Fig. \ref{fig:profiles_comparison}, note that the intersection between the SB profiles and the 30 $g'$ mag arcsec$^{-2}$ level (horizontal dotted line) spans a relatively narrow range of radii for TNG300 ($r^{1/4} \approx 4.5$--$4.9$ kpc$^{1/4}$ for circular apertures), but a much larger range for WWFI ($r^{1/4} \approx 3.8$--$4.9$ kpc$^{1/4}$ for circular apertures), which translates to larger variation in the `total' fluxes integrated out to the SB limit of 30 $g'$ mag arcsec$^{-2}$ and therefore larger variation in the ICL fractions (see equation \ref{eq:f_icl_corrected}). However, considering that the scatter in the SB profiles at smaller radii is roughly similar between simulation and observations, we believe that the larger scatter exhibited by observations at the 30 $g'$ mag arcsec$^{-2}$ level -- and therefore the larger scatter in the ICL fractions -- is mainly a consequence of observational limitations rather than `true' variation between clusters in the real Universe.

\subsection{Excess light above a de Vaucouleurs profile}
\label{subsec:f_icl_dev}

\begin{figure*}
  \centering
  \centerline{\hbox{
    \includegraphics[width=6cm]{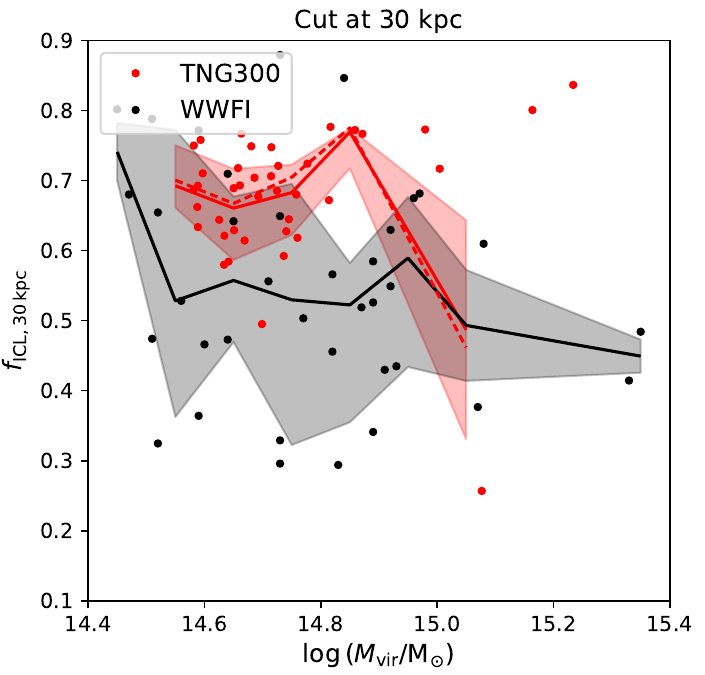}
    \includegraphics[width=6cm]{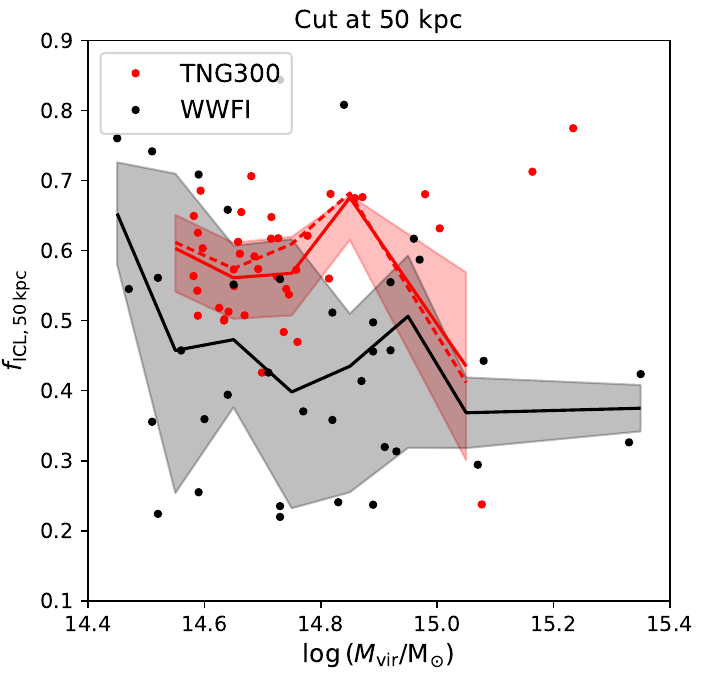}
    \includegraphics[width=6cm]{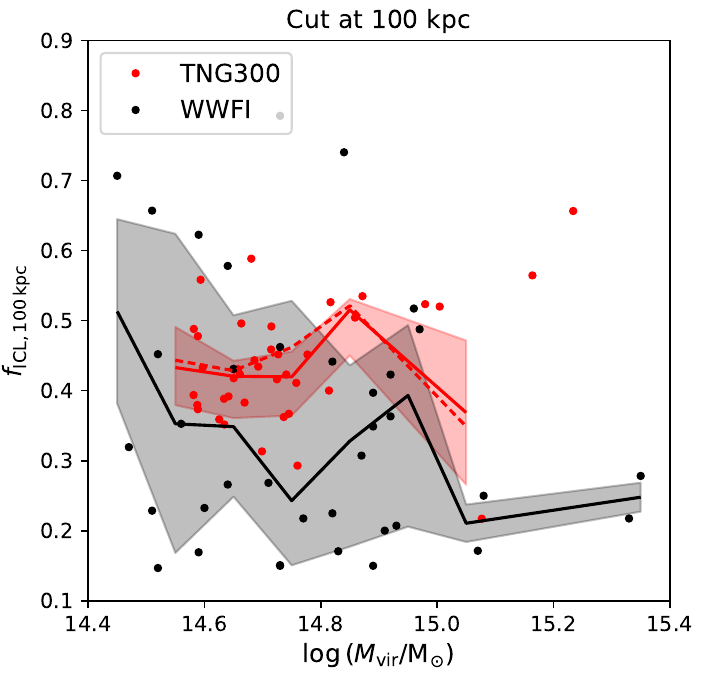}
  }}
  \caption{The ICL fraction for TNG300 (red) and WWFI (black), where the ICL consists of the light beyond a fixed circular aperture of radius 30 kpc (left), 50 kpc (middle), and 100 kpc (right), shown as a function of cluster mass. The dots, solid and dashed lines, and shaded regions have the same meaning as in previous figures. Only observed clusters with gravitational mass estimates (38 out of 170) are shown.}
	\label{fig:f_icl_fixed_kpc}
\end{figure*}

Another widely used ICL definition that was also included in the study by \cite{kluge2021photometric} consists in fitting a de Vaucouleurs profile \citep{deVaucouleurs1948} to the inner region of the BCG+ICL and then defining the ICL as the excess light over the outward extrapolation of the fitted model. As with the SB limit method from the previous section, we highlight the main methodological differences between our implementation and the original one by \cite{kluge2021photometric}:

\begin{itemize}
    \item[(i)] \cite{kluge2021photometric} fitted their models to the inner region brighter than 23 $g'$ mag arcsec$^{-2}$. While this works reasonably well for the WWFI observations, we find that the 23 $g'$ mag arcsec$^{-2}$ threshold is too close to the resolution limit of TNG300 (see Fig. \ref{fig:profiles_comparison}), resulting in too much weight being assigned to spurious regions of the simulated SB profiles. While such regions could be removed from the fitting procedure, this would leave a very small radial range for the de Vaucouleurs fit. Therefore, we instead relax the SB threshold so that intermediate regions of the BCG+ICL profile receive more weight. For consistency with the SB limit method from Section \ref{subsec:f_icl_sb27}, we fit the de Vaucouleurs models to the inner region brighter than 27 $g'$ mag arcsec$^{-2}$.

    \vspace{0.2cm}

    \item[(ii)] We perform all calculations directly on the 2D images rather than 1D profiles. To this end, we construct a smooth 27 $g'$ mag arcsec$^{-2}$ contour by applying the same procedure from the SB limit method (see Section \ref{subsec:f_icl_sb27}), then fit a de Vaucouleurs model to the region inside the contour using Astropy's Levenberg-Marquardt implementation, and finally calculate the excess light outside of the 27 $g'$ mag arcsec$^{-2}$ contour relative to the extrapolated model. The outer boundary is an appropriately orientated ellipse at the SB limit of 30 $g'$ mag arcsec$^{-2}$ (see Sections \ref{subsec:sersic_fits}--\ref{subsec:bcg+icl_profiles}). The flux correction for elliptical apertures (see Section \ref{subsec:undetected_light}) is applied to equation (\ref{eq:f_icl_corrected}).

\end{itemize}

The right-hand panel of Fig. \ref{fig:f_icl_sblim_27_and_dev} shows the ICL fractions obtained with this method for the TNG300 simulation (red) and WWFI observations (black), plotted against cluster mass. The median and scatter of the two samples are $f_{\rm ICL, deV} = 0.28 \pm 0.05$ for the simulation and $f_{\rm ICL, deV} = 0.26 \pm 0.19$ for the observations. The latter is significantly lower than the value of $0.48 \pm 0.20$ reported by \cite{kluge2021photometric}, which is mostly explained by the larger region used to carry out the fits in this work. Clearly, there is some arbitrariness in choosing the region that is used to fit the model, which we perceive as a weakness of this method (apart from the assumption that the BCG component is well described by a de Vaucouleurs profile). However, our ICL fraction measurements are closer to the values of $0.33 \pm 0.06$ and $0.21 \pm 0.12$ obtained by \cite{zibetti2005} and \cite{kluge2021photometric}, respectively, by fitting the de Vaucouleurs model over a larger radial interval ($\sim$15--80 kpc).

As seen from Fig. \ref{fig:f_icl_sblim_27_and_dev}, the agreement between the median values of $f_{\rm ICL, deV}$ obtained from the simulation and observations is excellent. The same cannot be said of the scatter, which is about 4 times larger in the observations, which we also attribute to observational limitations near the 30 $g'$ mag arcsec$^{-2}$ level, as explained in Section \ref{subsec:f_icl_sb27}, rather than to intrinsic scatter in the observed BCG+ICL systems.

\subsection{Twice the half-light radius}
\label{subsec:f_icl_2rhalf}

\begin{figure*}
  \centering
    \includegraphics[width=16.5cm]{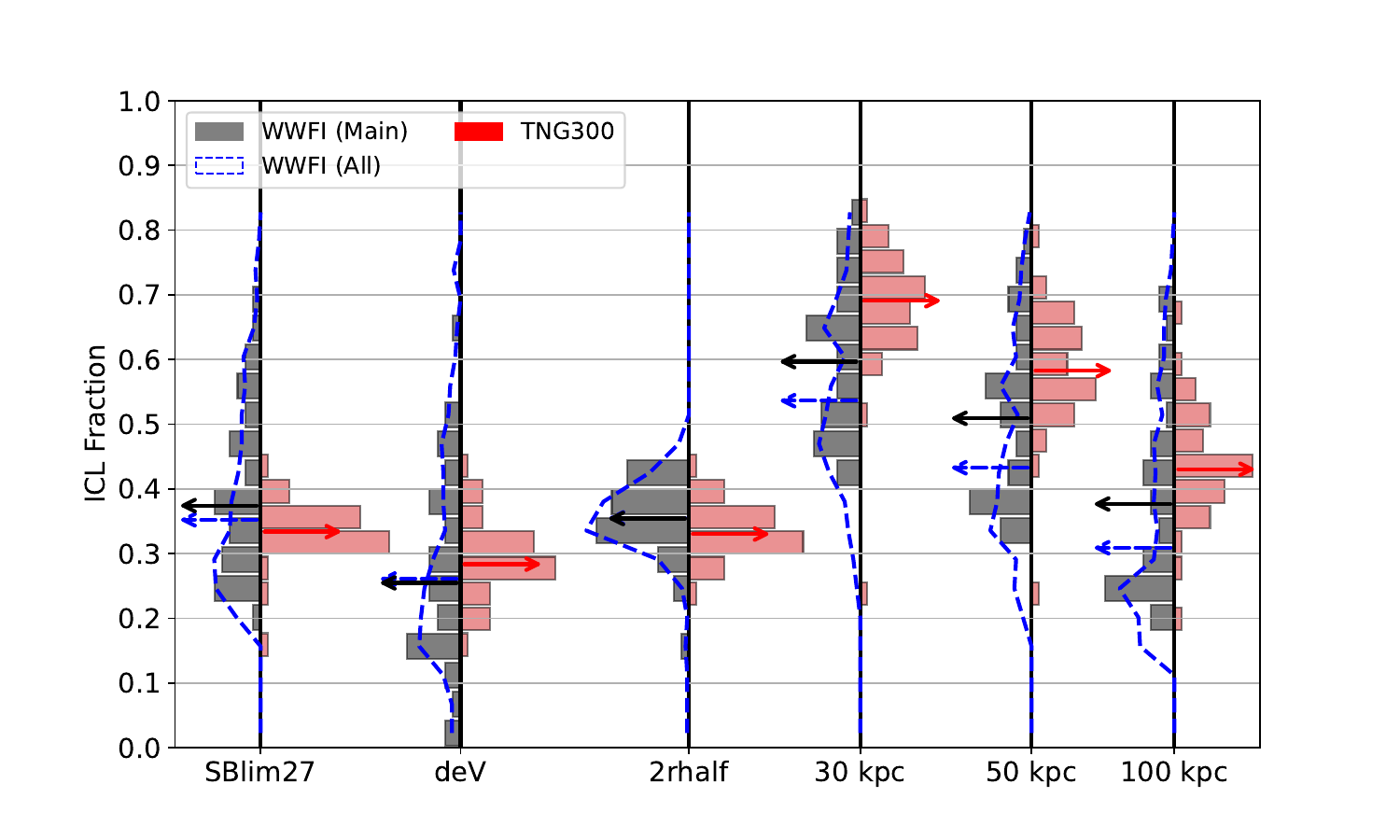}
  \caption{ICL fraction distributions for the various methods presented in Sections \ref{subsec:f_icl_sb27}--\ref{subsec:f_icl_fixed_kpc}. The black and red bars correspond to the observational (main WWFI sample, composed of 38 clusters) and simulated (TNG300, 40 clusters) cluster populations, respectively. The dashed blue lines show the distributions for the full WWFI sample, i.e. including all 170 observed clusters. The arrows indicate the medians of the distributions.}
	\label{fig:f_icl_barchart}
\end{figure*}

In simulations of galaxy formation, an aperture equivalent to twice the stellar half-mass radius has long been used as a practical way of separating galaxies from their surrounding stellar haloes \citep[e.g.][]{genel2014_Illustris, Pillepich2018a}. Although less common in ICL studies, this convention has recently been adopted in a series of simulation papers \citep[][]{proctor2024, montenegro2025}.

In observations of galaxy clusters, obtaining an accurate determination of the half-light radius of the BCG+ICL is complicated by the fact that its total light is not directly measurable. One solution to this problem, as done by \cite{kluge2021photometric}, is to extrapolate a Sérsic model fitted to the visible region of the BCG+ICL in order to estimate the total amount of light of the system. A different solution, which we implement in this work, is to use simulations to directly measure the total amount of light of the BCG+ICL out to $r_{\rm 200, crit}$, and compare it to the amount of light within the observational SB limit (30 $g'$ mag arcsec$^{-2}$) in order to obtain effective `correction' factors, as explained in Section \ref{subsec:undetected_light}.

We make use of the correction factors from equations (\ref{eq:correction_rhalf_circ})--(\ref{eq:correction_flux_ellip}) in order to obtain observationally feasible estimates of the half-light radius ($r_{\rm half}$) and the total flux of the system ($F_{\rm BCG+ICL}$), both for circular and elliptical apertures. Then we define the ICL as consisting of the light outside of $2 r_{\rm half}$ and compute the ICL fraction. As with the other methods, we perform all calculations directly on the 2D images with the help of the \texttt{photutils} package.

Fig. \ref{fig:f_icl_2rhalf} shows the ICL fraction obtained for TNG300 (red) and WWFI (black) by defining the ICL as the light outside of $2 r_{\rm half}$, for both circular (left) and elliptical (right) apertures, as a function of cluster mass. As before, the solid lines show the median results for the corrected, `SB-limited' measurements carried out on the `sigma-clipped' images (see Section \ref{subsec:background_subtraction}), while the dashed lines show the median results for the `full' measurements directly obtained from the `raw' synthetic images.

The median values and scatter obtained with this method are $f_{\rm ICL, 2rhalf} = 0.34 \pm 0.03$ ($f_{\rm ICL, 2rhalf} = 0.33 \pm 0.03$) for TNG300 and $f_{\rm ICL, 2rhalf} = 0.34 \pm 0.07$ ($f_{\rm ICL, 2rhalf} = 0.35 \pm 0.07$) for WWFI when using circular (elliptical) apertures. These values are not only consistent with each other, but also in very good agreement with the 3D stellar mass-based measurements from \cite{montenegro2025}. In that paper it was discussed that $2 r_{\rm half}$ seems to characterise the transition between the zone of gravitational influence of the BCG (including its stellar halo) and the diffuse medium of stars floating in the gravitational potential of the whole cluster (see also \citealt{chen2022} and \citealt{contini2024}).

\subsection{Fixed aperture}
\label{subsec:f_icl_fixed_kpc}

Finally, we consider the most practical definition of the ICL fraction, which is to simply impose a circular aperture of fixed radius and assign all of the exterior light (excluding satellites) to the ICL. In particular, we consider radii of 30, 50, and 100 kpc, which are common choices in the literature.

Fig. \ref{fig:f_icl_fixed_kpc} shows our measurements of the ICL fraction in TNG300 (red) and WWFI (black) for circular apertures of 30 kpc (left), 50 kpc (middle), and 100 kpc (right). As expected, the ICL fractions become smaller as the radial cut increases from left to right. The median ICL fraction and associated scatter we obtain for the 30, 50 and 100 kpc apertures in TNG300 are $0.69 \pm 0.07$, $0.58 \pm 0.08$ and $0.43 \pm 0.08$, respectively, and $0.53 \pm 0.16$, $0.43 \pm 0.18$ and $0.29 \pm 0.18$ for the WWFI observations.

Unlike the other methods, which show good agreement in the median ICL fraction between the simulation and observations, the median ICL fractions for the fixed aperture method tend to be larger in TNG300 relative to the WWFI observations, which is a direct consequence of the simulated SB profiles being systematically larger than the observed ones at any fixed radius (see Fig. \ref{fig:profiles_comparison}). We note, however, that in the case of the observations, the ICL fractions obtained for the 100 kpc aperture are comparable to those obtained with the other methods, suggesting that a fixed aperture of 100 kpc might represent a practical boundary between the BCG and ICL that is easy to implement, at least in the mass range explored here. It is also reasonably close to the median `observationally derived' transition radius of 128 kpc obtained by \cite{Brough+2024}.

\subsection{Overview}
\label{subsec:f_icl_overview}

Fig. \ref{fig:f_icl_barchart} shows an overview of all the ICL fraction definitions considered so far, once again comparing simulation (red) versus observations (black). In the case of the $2 \, r_{\rm half}$ method, we only show the results for elliptical apertures, given that the results for circular apertures are extremely similar, as seen in Section \ref{subsec:f_icl_2rhalf}. Fig. \ref{fig:f_icl_barchart} includes measurements for all the 170 observed clusters, but we note that restricting the sample to only those observed clusters with gravitational mass estimates would yield very similar results. Furthermore, we note that the ICL fractions for the different methods presented in Sections \ref{subsec:f_icl_sb27}--\ref{subsec:f_icl_fixed_kpc} do not show a strong dependence on cluster mass, which means that Fig. \ref{fig:f_icl_barchart} captures the main trends from Figs \ref{fig:f_icl_sblim_27_and_dev}--\ref{fig:f_icl_fixed_kpc}.

We can see from Fig. \ref{fig:f_icl_barchart} that the \textit{median} ICL fractions are in good agreement between the simulation and observations for almost all methods, with the exception of the fixed aperture measurements (for which the TNG300 ICL fractions show an excess of about 15 per cent). However, the \textit{scatter} in $f_{\rm ICL}$ is always larger for the observations than the simulation, which we have primarily attributed to observational limitations near the SB limit of 30 $g'$ mag arcsec$^{-2}$ rather than intrinsic variation in the observed systems.

\section{Summary and discussion}
\label{sec:summary_and_discussion}

We have conducted a robust, `apples-to-apples' comparison between the photometric properties of the ICL in simulations and observations. This was achieved by generating $g'$-band synthetic images of 40 massive ($\log\left(M_{200}/{\rm M}_{\odot}\right) > 14.5$) galaxy clusters from the TNG300 simulation via a procedure that closely matches WWFI observations of 170 clusters at $z \approx 0.06$ \citep{Kluge+2020}, and then applying the same satellite-masking procedure to the synthetic images that was used for the observations. The main steps of the synthetic image generation procedure are illustrated in Fig. \ref{fig:cluster_multipanel}. Furthermore, for the sake of consistency, we applied the same background subtraction procedure to the synthetic and real images (Section \ref{subsec:background_subtraction}). Some examples from both image data sets, with and without masking, can be found in Appendix \ref{app:cluster_images}.

By analyzing the BCG+ICL as a single entity (i.e. without attempting to separate the BCG from the ICL), we found from our 2D image analysis that the simulated systems have higher ellipticities ($0.35 \pm 0.14$ versus $0.19 \pm 0.10$) and higher Sérsic indices ($6.1 \pm 2.4$ versus $4.3 \pm 1.4$) than the observed ones (Fig. \ref{fig:morph_vs_mvir_comparison}). Similarly, the BCG+ICL light distributions are systematically more extended (by about a factor of 2) and brighter (by about 1 $g'$ mag arcsec$^{-2}$) in the simulation than in the observations, as can be seen in Figs \ref{fig:profiles_comparison}, \ref{fig:rpetro} and \ref{fig:rhalf}. Typical values of BCG+ICL half-light radii are in the order of $\sim$30 kpc ($\pm$0.4 dex) for the observations and $\sim$70 kpc ($\pm$0.2 dex) for the simulations.

The discrepancies in BCG+ICL sizes between TNG300 and observations are not entirely surprising, since previous works had found similar results at the massive end of the galaxy population. In particular, \cite{genel2018size} found that the $r$-band projected sizes of massive, quenched galaxies from the TNG100 simulation were about 0.2 dex larger than their observational counterparts. Furthermore, using forward-modeling techniques similar to those used in this work, \cite{Rodriguez-Gomez2019} also found that the sizes of massive TNG100 galaxies were about 0.2 dex larger than those from Pan-STARRS observations (see their fig. 7). The differences between theoretical and observed sizes reported in the current work are somewhat larger ($\gtrsim$ 0.3 dex) than those from previous studies, which could be due to a combination of factors, such as (i) increased emphasis on the massive end of the galaxy population, well beyond the largest galaxies produced in the smaller TNG100 simulation, (ii) decreased realism due to resolution effects, since the original IllustrisTNG model was developed and tested at TNG100 resolution (at 8 times better mass resolution than TNG300), and (iii) use of the $g'$ band, which is generally not a good tracer of stellar mass and therefore is `further away' from the observables considered during the model design (see Section \ref{subsec:simulation}).

Similarly, the fact that the simulated SB profiles are systematically brighter than the observed ones is not entirely unexpected. \cite{ardila2021} compared the stellar surface density profiles of galaxies from the TNG100 simulation to those from Hyper Suprime-Cam observations over a wide range of halo masses (see their fig. 5). While they found overall good agreement between TNG100 and observations, they also found that the amplitudes of the stellar surface density profiles tend to grow faster with halo mass in the simulation than in observations, leading to a clear systematic offset in their most massive bin ($\log_{10}\left(M_{\rm vir} / {\rm M}_{\odot} \right) = 13.83$--$14.25$). In addition, \cite{montenegro2023} found that TNG300 clusters exhibit a slight stellar mass excess at the extremely massive end of the stellar-to-halo mass relation (see their fig. 1) compared to observational estimates by \cite{DeMaio2020}. \cite{Pillepich2018a} and \cite{montenegro2025} also showed that the BCG+ICL stellar masses of the most massive clusters appear to be larger than those fitted to the observations in \cite{Kravtsov2018}. It is possible that this slight excess in the BCG+ICL stellar mass might become magnified in the $g'$ band, which is not a reliable tracer of stellar mass (compared to redder bands such as $r'$ and $i'$).

The fact that our simulated BCG+ICL systems are brighter than their observational counterparts can provide clues about some aspects of the IllustrisTNG galaxy formation model that might require improvement. As noted in the previous paragraph, the excess in BCG+ICL luminosity (or stellar mass) relative to observations only manifests at the massive end of the cluster population ($\log_{10}\left(M_{\rm vir} / {\rm M}_{\odot} \right) \gtrsim 14$). While it might be tempting to attribute this discrepancy to implementation details of the `kinetic' mode of the AGN feedback model \citep{Weinberger2017} in which these BCGs operate at the present time, it is important to note that these systems quenched billions of years ago ($z \gtrsim 1$) and that over 80 per cent of their stellar mass content consists of \textit{ex situ} (i.e. accreted) stars \citep{Pillepich2018a, montenegro2023}. Furthermore, \cite{Pillepich2018a} found that 90 per cent of the accreted stellar mass in clusters with $M_{\rm 200, crit} \gtrsim 10^{14} \, \Msun$ was acquired from galaxies with stellar masses of at least $3$--$5 \times 10^{10} \, \Msun$, on average, which is not too dissimilar from the current stellar mass of the Milky Way. Therefore, what really matters could possibly be the regulation of star formation in these galaxies, and at somewhat earlier times, rather than in the massive systems as we see them today. Nevertheless, since the focus of this work was on \textit{relative} measurements of the ICL, particularly ICL fractions, such offsets in the SB profiles do not necessarily translate to differences in our other measurements.

In fact, the main goal of this work was to provide a consistent comparison of the ICL fraction ($f_{\rm ICL}$), defined as the ratio between the light components associated with the ICL and the BCG+ICL, between the TNG300 simulation and WWFI observations. To this end, we considered four different ICL definitions: (i) the light dimmer than a fixed SB threshold of 27 $g'$ mag arcsec$^{-2}$ (Section \ref{subsec:f_icl_sb27}), (ii) the excess light over a de Vaucouleurs model fitted to the inner region brighter than 27 $g'$ mag arcsec$^{-2}$ (Section \ref{subsec:f_icl_dev}), (iii) the light beyond twice the half-light radius (Section \ref{subsec:f_icl_2rhalf}), and (iv) the light beyond a fixed circular aperture of radius 30, 50 or 100 kpc (Section \ref{subsec:f_icl_fixed_kpc}). In all cases, we performed the necessary measurements directly on the 2D images -- real or synthetic -- using the same photometric analysis pipeline.

Our first finding regarding the ICL fractions is that they are not strongly dependent on cluster mass, at least for the mass range considered (Figs \ref{fig:f_icl_sblim_27_and_dev}--\ref{fig:f_icl_fixed_kpc}; see also \citealt{Pillepich2018a}). This justifies calculating the overall $f_{\rm ICL}$ distributions and combining them in Fig. \ref{fig:f_icl_barchart}, which represents an overview of how the ICL fractions compare between simulation and observations for the four different methods. The two main takeaways from this plot are the following:

\begin{itemize}

    \item[(i)] The median values of $f_{\rm ICL}$ in simulation and observations are mostly consistent with each other, with the exception of the fixed-aperture method, for which the simulated clusters exhibit $\sim$15 per cent more ICL than the observed ones (this is consistent with the larger half-light radii exhibited by the simulated systems, which indicates that the systems are more extended and therefore contain a larger fraction of the BCG+ICL light beyond any fixed radius).

    \vspace{0.2cm}

    \item[(ii)] The scatter in $f_{\rm ICL}$ is larger in the observations than in the simulation. This is mainly a consequence of uncertainties in observational determinations of the total BCG+ICL light, as can be seen from the `horizontal' scatter in the observed SB profiles near the SB limit of 30 $g'$ mag arcsec$^{-2}$ in Fig. \ref{fig:profiles_comparison}.

\end{itemize}

While a detailed discussion of which of the various ICL definitions considered is the most meaningful or physically motivated would be beyond the scope of this work, we note that the median $f_{\rm ICL}$ for most of the methods lies in the vicinity of 0.3 (except for the 30 and 50 kpc definitions, which yield larger values). This is in good agreement with the ICL fraction measurements from \cite{montenegro2025} obtained by applying the $2 r_{\rm half}$ method to the 3D stellar mass distributions of TNG300 clusters, while also being roughly consistent with a BCG/ICL transition radius around 100 kpc, in broad agreement with \cite{Brough+2024}. This also suggests that most of the definitions considered here, despite their methodological differences, are motivated by common features of the BCG+ICL light distribution, which in turn are a reflection of underlying dynamical properties and other physical characteristics of galaxy clusters \citep[][]{rudick2011, cui2014, Brough+2024}. Such connections between the kinematic and photometric properties of galaxy clusters will be explored further in upcoming work.

Next-generation galaxy formation models will benefit from comparisons such as the ones we have presented in this paper, since they can be used to assess the numerical implementation of astrophysical processes, such as AGN feedback, across a wide range of galaxy mergers in more extreme environments. The need for such enhanced comparisons between theory and observations will grow as galaxy formation models continue to evolve, and as upcoming surveys -- such as the Rubin Observatory `Legacy Survey of Space and Time' \citep{robertson2019_LSST} -- produce increasingly large, deep, and detailed data sets. Such advances on the theoretical and observational fronts, combined with forward-modeling techniques such as the one presented in this work, will significantly expand our understanding of the ICL in the coming years.

\vspace{0.5cm}

\section*{Acknowledgements}

% The Acknowledgements section is not numbered. Here you can thank helpful
% colleagues, acknowledge funding agencies, telescopes and facilities used etc.

We thank Laura Sales and Eric Rohr for useful comments and discussions, as well as the referee for an insightful report. DMT thanks SECIHTI (previously CONAHCyT) for a PhD fellowship. DMT, VAR and AM acknowledge financial support from the CONAHCyT grant CF G-543 and the DGAPA-PAPIIT grant IN106823. DMT and BCS acknowledge financial support from the DGAPA-PAPIIT grant IN111825 and IN108323. AM acknowledges  Universidad Nacional Autónoma de México Postdoctoral Program (POSDOC) for financial support. The Wendelstein 2.1 m telescope project was funded by the Bavarian government and by the German Federal government through a common funding process. Part of the 2.1 m instrumentation including some of the upgrades for the infrastructure were funded by the Cluster of Excellence “Origin of the Universe” of the German Science foundation DFG. The IllustrisTNG flagship simulations were run on the HazelHen Cray XC40 supercomputer at the High Performance Computing Center Stuttgart (HLRS) as part of project GCS-ILLU of the Gauss Centre for Supercomputing (GCS). Ancillary and test runs of the project were also run on the compute cluster operated by HITS, on the Stampede supercomputer at TACC/XSEDE (allocation AST140063), at the Hydra and Draco supercomputers at the Max Planck Computing and Data Facility (MPCDF), and on the MIT/Harvard computing facilities supported by FAS and MIT MKI. Some calculations were performed on the VERA supercomputer at the MPCDF.

%%%%%%%%%%%%%%%%%%%%%%%%%%%%%%%%%%%%%%%%%%%%%%%%%%
\section*{Data Availability}

% The inclusion of a Data Availability Statement is a requirement for articles published in MNRAS. Data Availability Statements provide a standardised format for readers to understand the availability of data underlying the research results described in the article. The statement may refer to original data generated in the course of the study or to third-party data analysed in the article. The statement should describe and provide means of access, where possible, by linking to the data or providing the required accession numbers for the relevant databases or DOIs.

The data from the IllustrisTNG simulation used in this work are publicly available at the website \href{https://www.tng-project.org}{https://www.tng-project.org} \citep{nelson2019illustristng}.



\bibliographystyle{mnras}
\bibliography{paper}

%\begin{thebibliography}{99}
%\end{thebibliography}

%%%%%%%%%%%%%%%%%%%%%%%%%%%%%%%%%%%%%%%%%%%%%%%%%%

%%%%%%%%%%%%%%%%% APPENDICES %%%%%%%%%%%%%%%%%%%%%

\appendix

\section{Cluster images}
\label{app:cluster_images}

\begin{figure*}
%\centering
\centerline{\vbox{
    \includegraphics[width=\linewidth]{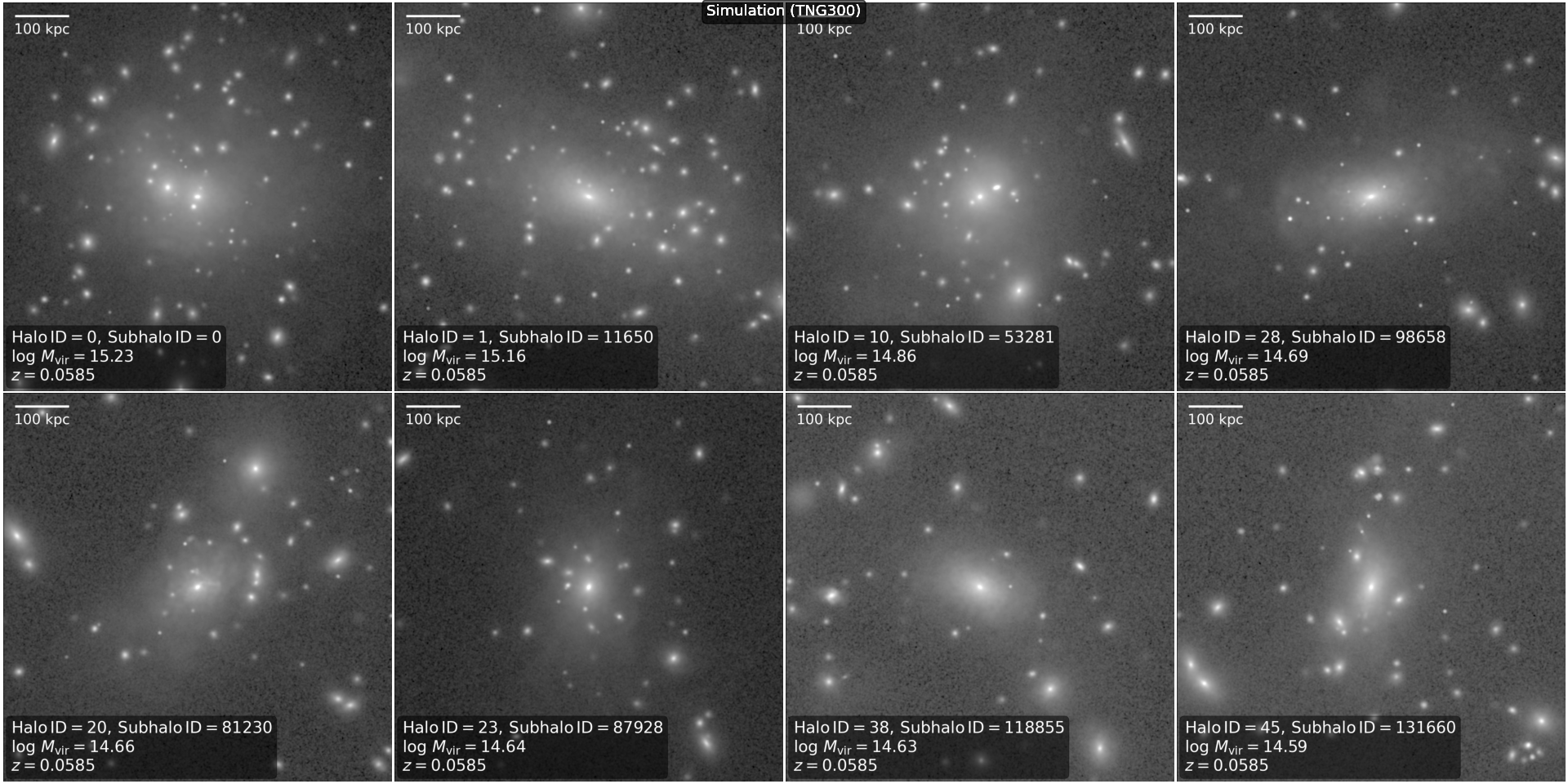}
    \includegraphics[width=\linewidth]
    {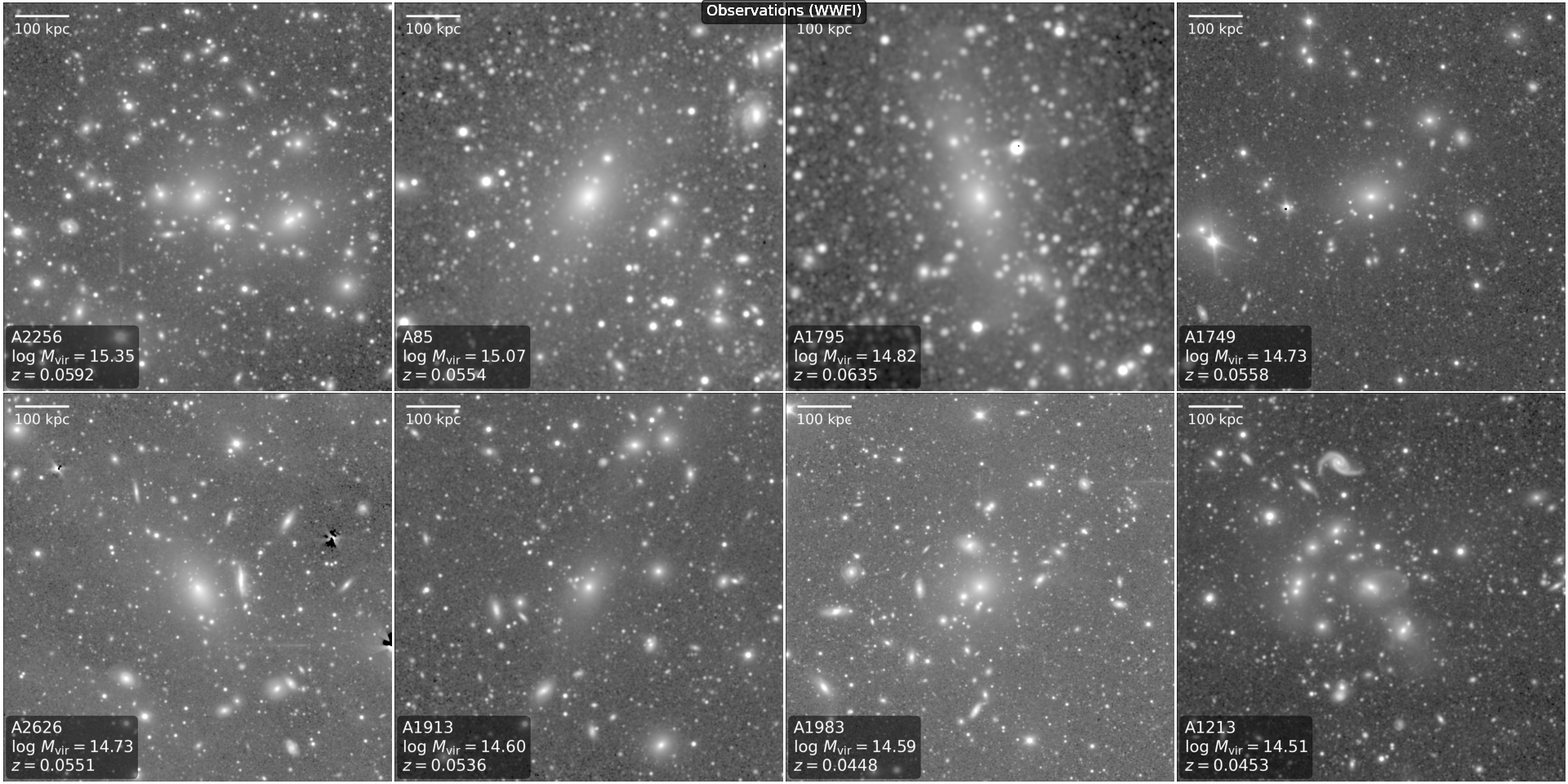}  
  }}
\caption{Randomly chosen examples of \textit{synthetic} images of TNG300 clusters (upper two rows) and \textit{real} images of WWFI clusters (bottom two rows), shown in order of decreasing virial mass. The field of view of each image is 750 kpc. (Note that we use a larger region for our analysis.) Clearly, the real images contain a large amount of Galactic stars, as well as some foreground and background (mostly the latter) galaxies. All of these `contaminants', along with the cluster galaxies (i.e. satellites), are masked in order to analyze the BCG+ICL.}
\label{fig:cluster_images}
\end{figure*}

\begin{figure*}
%\centering
%\includegraphics[width=\linewidth]{fig/collage_real_mock_plus_mask_4x4_750kpc.png}
\centerline{\vbox{
    \includegraphics[width=\linewidth]{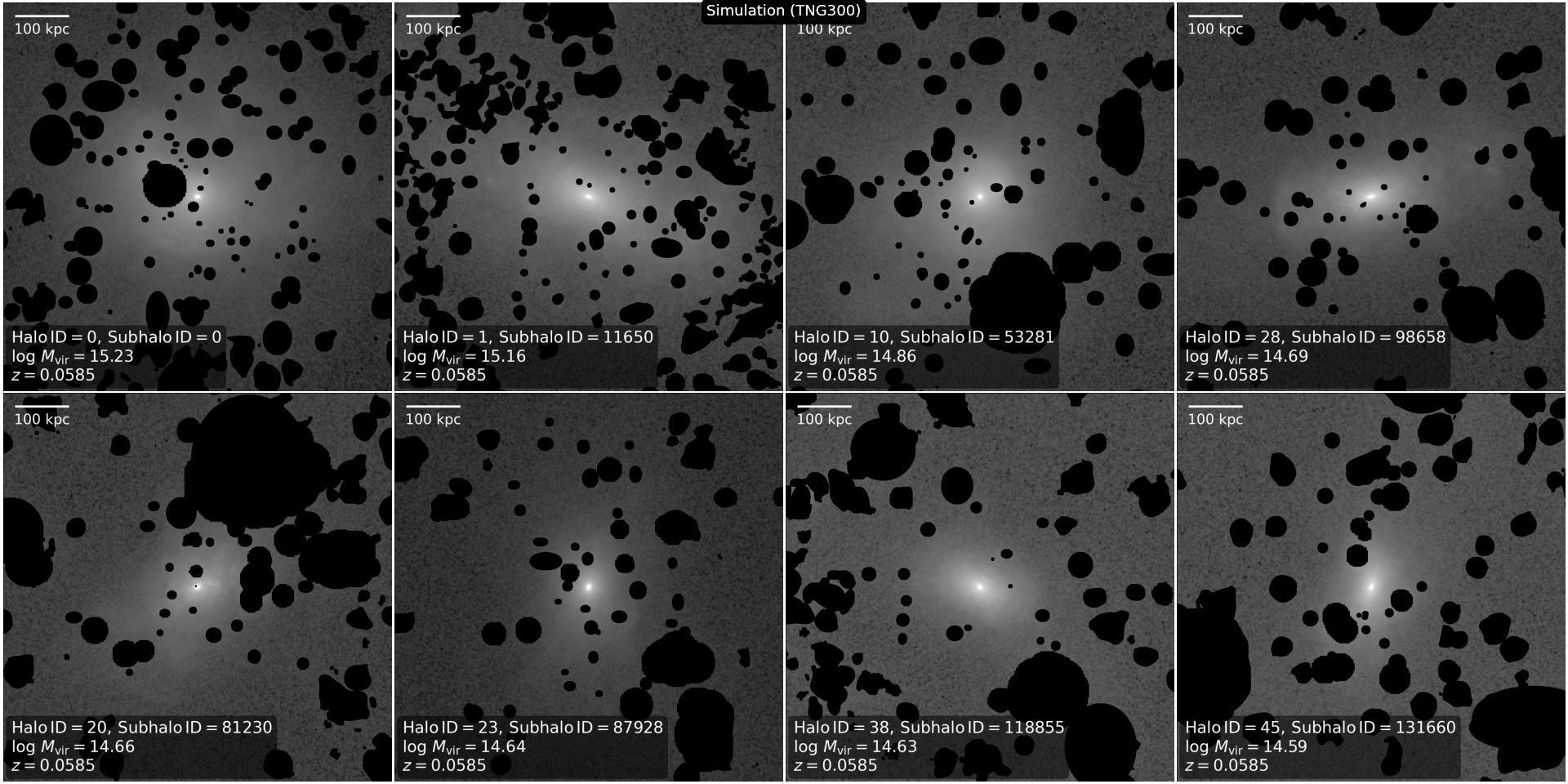}
    \includegraphics[width=\linewidth]{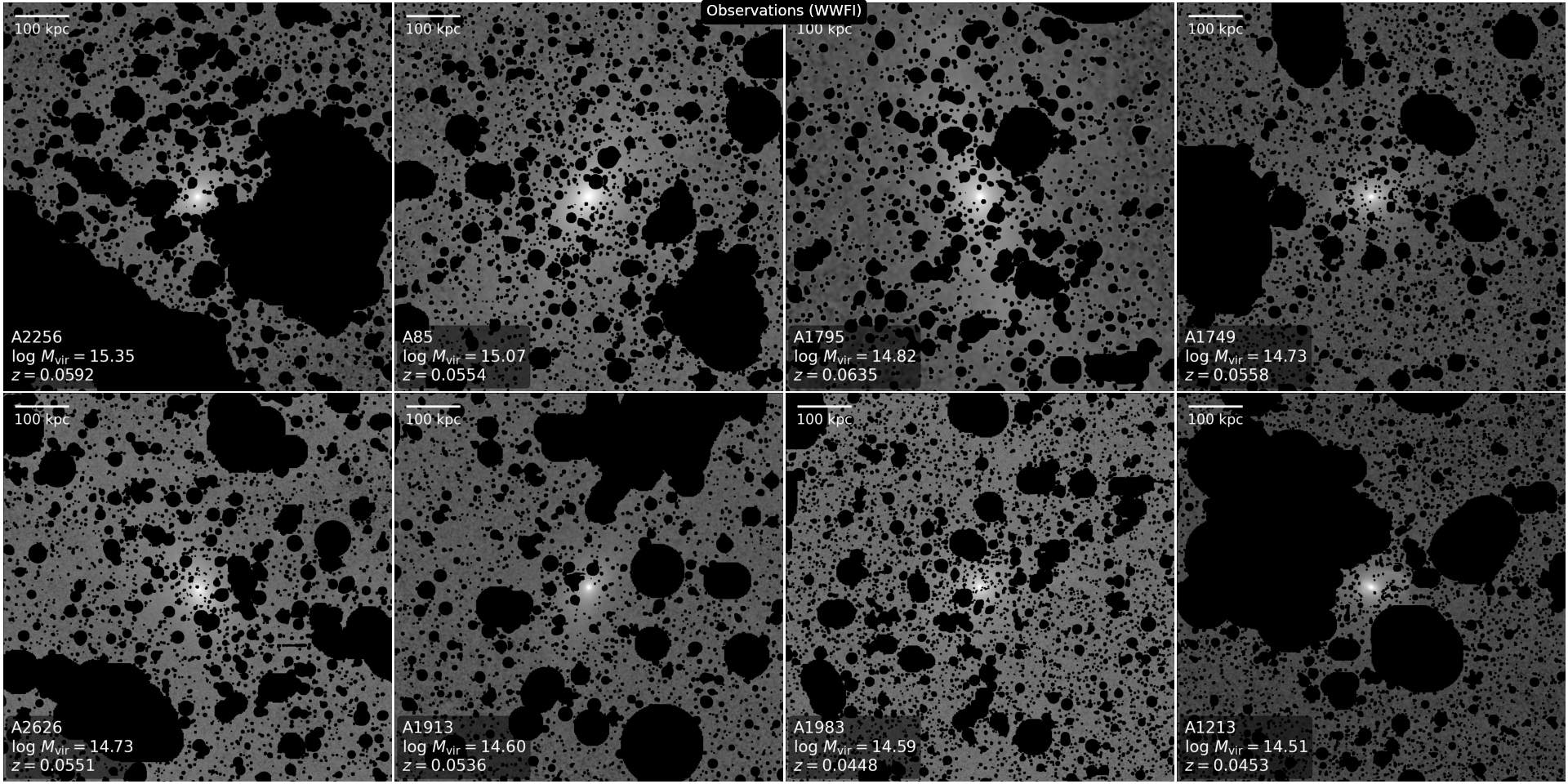}  
  }}
\caption{Masked version of the cluster images from Fig. \ref{fig:cluster_images}. The non-masked light is assumed to belong to the BCG+ICL.}
\label{fig:cluster_images_masked}
\end{figure*}

Fig. \ref{fig:cluster_images} shows randomly chosen examples of the synthetic (upper two rows) and real (bottom two rows) cluster images used in this work (see Section \ref{subsec:synthetic_images}), ordered by decreasing cluster mass. Fig. \ref{fig:cluster_images_masked} shows the same objects with the satellite mask applied (see Section \ref{subsec:image_masking}).

\section{Impact of image smoothing on ICL fractions}
\label{app:image_smoothing}

As discussed in \cite{Tang2018}, the amount of image smoothing can have a non-negligible impact on ICL fraction measurements. In particular, excessive smoothing can `spread' light from the BCG and satellites into the ICL, resulting in overestimated ICL fractions.  In order to assess whether the 64-neighbour ($N_{\rm ngb} = 64$) adaptive smoothing used in this work (see Section \ref{subsec:synthetic_images}) is sufficiently accurate for our purposes, we have repeated our analysis using synthetic images generated with different levels of smoothing, specifically for $N_{\rm ngb} = $ 16, 32, 128 and 256.

\begin{figure}
  \centering
  \includegraphics[width=8.5cm]{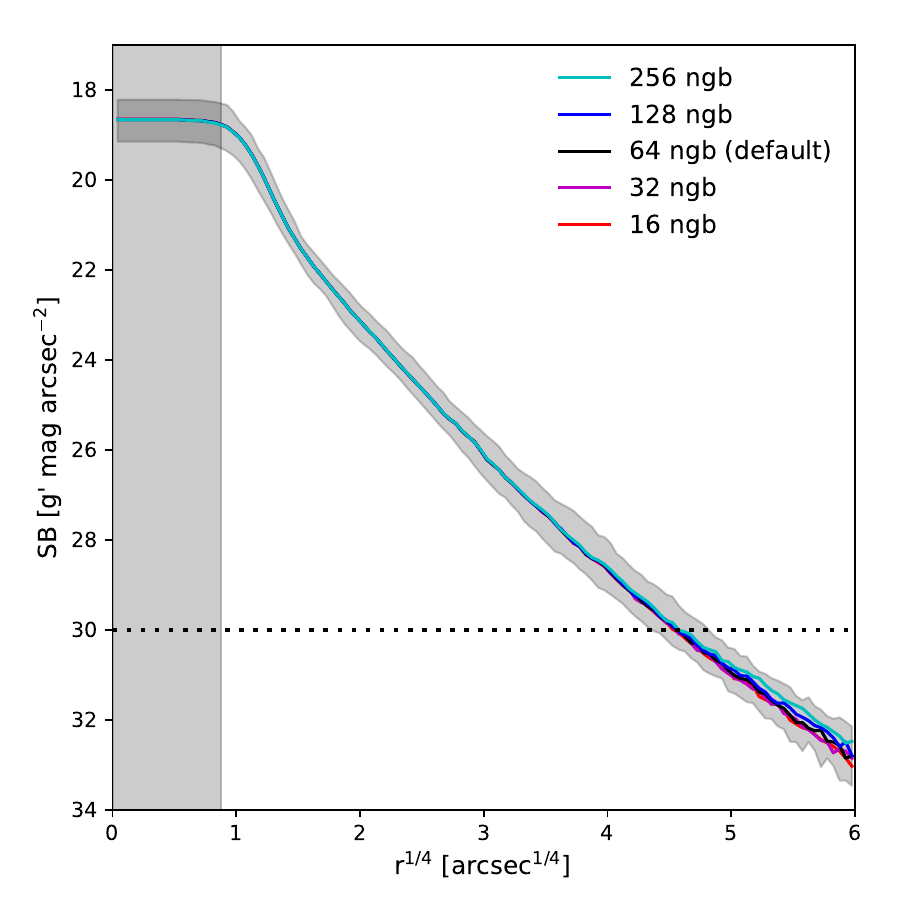}
	\caption{Median SB profiles of the \textit{raw} synthetic images (i.e. before applying the sigma-clipping procedure described in Section \ref{subsec:background_subtraction}), where the different colours correspond to images generated with different levels of adaptive smoothing of the stellar particles (see Section \ref{subsec:synthetic_images}). The grey shaded region around the SB profiles represents the 16th to 84th percentile range for the images generated with 64-neighbour smoothing (the default), whereas the grey shaded region to the left indicates the resolution limit of the observations (0.5 times the PSF FWHM). The horizontal dotted line represents the observational SB detection limit (30 $g'$ mag arcsec$^{-2}$).}
	\label{fig:sb_profiles_ngb}
\end{figure}

\begin{figure}
  \centering
  \includegraphics[width=8.5cm]{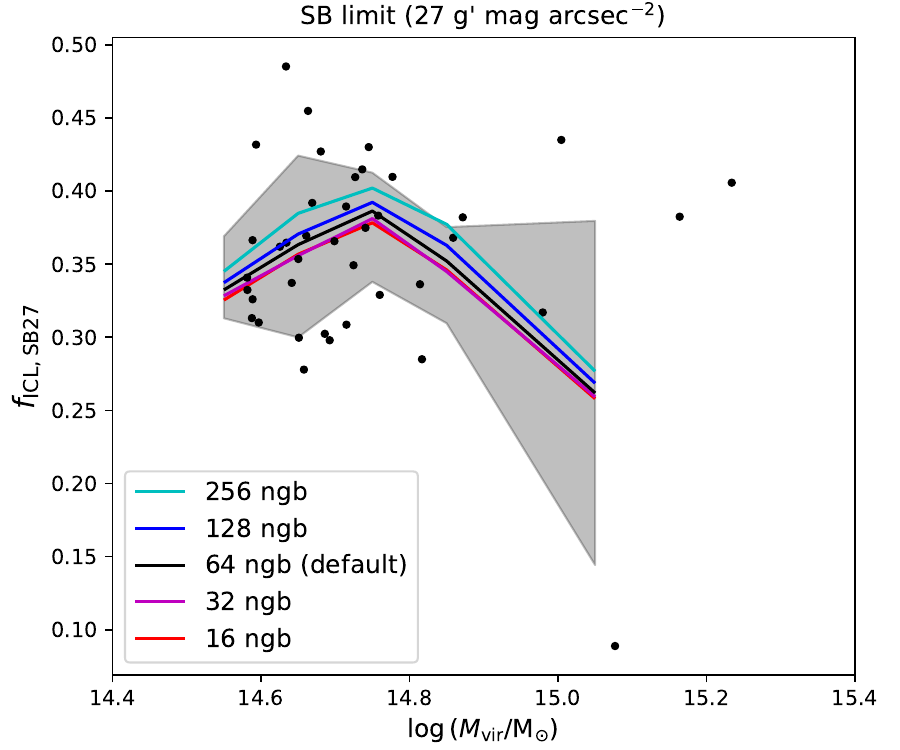}
	\caption{Median ICL fractions obtained for the \textit{raw} synthetic images (see Section \ref{subsec:background_subtraction}) using the SB limit method (see Section \ref{subsec:f_icl_sb27}), plotted as a function of cluster mass. As in Fig. \ref{fig:sb_profiles_ngb}, the different colours correspond to different levels of adaptive smoothing. The individual black points represent individual measurements for the images with 64-neighbour smoothing (the default), whereas the grey shaded region indicates the corresponding 16th to 84th percentile range. Note that no observationally motivated correction factors (see Section \ref{subsec:undetected_light}) were applied in this figure, i.e. all photometric measurements were carried out to $r_{\rm 200, crit}$.}
	\label{fig:f_icl_sblim_ngb}
\end{figure}

Fig. \ref{fig:sb_profiles_ngb} shows the impact of the different smoothing lengths on the \textit{raw} SB profiles (i.e. those obtained directly from the realistic synthetic images, without applying the background-subtraction procedure from Section \ref{subsec:background_subtraction}). Although the impact of smoothing is negligible at SB levels brighter than the observational SB detection limit of 30 $g'$ mag arcsec$^{-2}$, at larger radii the SB profiles become slightly brighter for higher levels of smoothing. While these differences are small in magnitude, and in fact would not be discernible within observational limits, they could potentially have an impact on \textit{theoretical} measurements of the ICL fraction. We explore this possibility in Fig. \ref{fig:f_icl_sblim_ngb}, which shows the effect of different smoothing lengths on the ICL fractions obtained from the \textit{raw} synthetic images using the SB limit method (see Section \ref{subsec:f_icl_sb27}). In agreement with the low-redshift results from \cite{Tang2018}, we find that higher levels of smoothing lead to larger ICL fractions, although the effect is relatively small.

More precisely, we find that the median (plus scatter) of the ICL fractions in Fig. \ref{fig:f_icl_sblim_ngb} at the default smoothing level ($N_{\rm ngb} = 64$, black) is $0.365 \pm 0.052$, whereas the corresponding value at the finest level of smoothing considered ($N_{\rm ngb} = 16$, red) is $0.358 \pm 0.050$. (Also note that the ICL fractions appear to be almost perfectly converged at the two finer levels of smoothing, $N_{\rm ngb} =$ 16 and 32, suggesting that the results would be similar for even finer levels of smoothing.) This represents a difference of $0.007$, or less than 1 per cent of the total BCG+ICL luminosity, which we consider to be sufficiently accurate for our purpose of comparing the ICL fraction between simulations and observations. The other ICL definitions considered in Section \ref{sec:icl_fraction} yield similarly small differences.

%%%%%%%%%%%%%%%%%%%%%%%%%%%%%%%%%%%%%%%%%%%%%%%%%%

% Don't change these lines
\bsp	% typesetting comment
\label{lastpage}
\end{document}